\documentclass[aps,prb,reprint,superscriptaddress,floatfix,nofootinbib]{revtex4-2}
\usepackage{graphicx}
\usepackage{amsmath}
\usepackage{bm}
\usepackage{xcolor}
\usepackage{verbatim}

\usepackage{multirow}

\newcommand{\bra}[1]{\left<#1\right|} 
\newcommand{\ket}[1]{\left|#1\right>} 
\newcommand{\braket}[2]{\left<#1|#2\right>}

\newcommand{\angstrom}{\mbox{\normalfont\AA}}

\begin{document}


\title{Pseudospin-electric coupling for holes beyond the envelope-function approximation}


\author{Pericles Philippopoulos}
\affiliation{Department of Physics, McGill University, 3600 rue University, Montreal, Qc H3A 2T8, Canada}

\author{Stefano Chesi}
\affiliation{Beijing Computational Science Research Center, Beijing 100193, China}
\affiliation{Department of Physics, Beijing Normal University, Beijing 100875, China}

\author{Dimitrie Culcer}
\affiliation{School of Physics, University of New South Wales, Kensington, NSW 2052, Australia}
\affiliation{Australian Research Council Centre of Excellence in Low-Energy Electronics Technologies, The University of New South Wales, Sydney 2052, Australia }

\author{W. A. Coish}
\affiliation{Department of Physics, McGill University, 3600 rue University, Montreal, Qc H3A 2T8, Canada}


\date{\today}



\begin{abstract}
In the envelope-function approximation, interband transitions produced by electric fields are neglected.
However, electric fields may lead to a spatially local ($k$-independent) coupling of band (internal, pseudospin) degrees of freedom.  
Such a coupling exists between heavy-hole and light-hole (pseudo-)spin states in III-V semiconductors, such as GaAs, or in group IV semiconductors (germanium, silicon, ...) with broken inversion symmetry. 
Here, we calculate the electric-dipole (pseudospin-electric) coupling for holes in GaAs from first principles.
We find a transition dipole of $0.5$ debye, a significant fraction of that for the hydrogen-atom $1s\to2p$ transition.
In addition, we derive the Dresselhaus spin-orbit coupling that is generated by this transition dipole for heavy holes in an asymmetric quantum well.
A quantitative microscopic description of this pseudospin-electric coupling may be important for understanding the origin of spin splitting in quantum wells, spin coherence/relaxation ($T_2^*/T_1$) times, spin-electric coupling for cavity-QED, electric-dipole spin resonance, and spin non-conserving tunneling in double quantum dot systems.  
\end{abstract}

\maketitle

\section{Introduction}\label{sec:Intro}
A standard theoretical tool for studying electrons (or holes) confined to nanostructures is $\bm{k} \cdot \bm{p}$ theory under the envelope function approximation \cite{winkler2003spin, voon2009kp}.
Within this formalism, one ignores interband coupling due to an electric field ($\propto \bm{\nabla}U$) arising from the slowly-varying part of the potential, $U$.
This established procedure has been used to understand two distinct forms of spin-orbit coupling for semiconductors \cite{winkler2003spin}: Rashba spin-orbit coupling \cite{bychkov1984oscillatory} (arising from inversion asymmetry in $U$, structure inversion asymmetry) and Dresselhaus spin-orbit coupling \cite{dresselhaus1955spin} (arising from inversion asymmetry in the underlying crystal, bulk inversion asymmetry).

Beyond the envelope function approximation, an electric field can lead to interband coupling. 
For states that transform like the $\Gamma_8$ representation of the tetrahedral double group ($T_d$), this coupling can be parameterized by a single constant, $\chi$ \cite{bir1963spin}.
The unit cell of, e.g., bulk silicon or germanium has a center of inversion symmetry,\footnote{Strictly speaking, silicon is not inversion symmetric. 
It has a diamond crystalline structure and is therefore symmetric under the (non-symmorphic) symmetry operation which can be described by an inversion of the crystal about a lattice site followed by a translation of the crystal by (1/4,1/4,1/4).
Although it is not a proper inversion, this symmetry still precludes interband electric-dipole coupling in the valence-band subspace.} which precludes any interband electric-dipole coupling in the valence-band subspace, leading to $\chi=0$.
However, in the presence of an acceptor (or any other impurity), inversion symmetry is broken and the point-group symmetry is reduced to $T_d$. 
This reduction in symmetry gives rise to a finite electric-dipole matrix element between the heavy-hole and light-hole states (which transform like the $\Gamma_8$ representation of $T_d$) \cite{bir1963spin}. 
The interband coupling influences the spectrum of acceptors in silicon in the presence of an electric field \cite{kopf1992linear} and could allow for better control of acceptor spin qubits in silicon \cite{abadillo2018entanglement}. 

In contrast to silicon or germanium, III-V semiconductors do not have a center of inversion symmetry and the valence-band states transform according to the $\Gamma_8$ representation of $T_d$.
Therefore, the electric field can couple valence-band states, even in the absence of an acceptor.  
Specifically, this interband coupling allows an AC electric field to generate Rabi oscillations between heavy-hole and light-hole pseudospin states, even without heavy-hole/light-hole mixing. 
In the case of III-V semiconductors, the interband coupling parameter $\chi$ is a material parameter that can be evaluated given the crystal eigenstates (Bloch functions).   
Recently, the Bloch amplitudes for the valence band of GaAs have been approximated using the Kohn-Sham orbitals from an all-electron density-functional theory calculation \cite{philippopoulos2020first}.
Here, we evaluate the $\bm{k}\cdot\bm{p}$ parameters and band gaps in GaAs using the technique described in Ref.~\cite{philippopoulos2020first}.
The resulting calculated parameters are in reasonable agreement with empirically established $\bm{k}\cdot\bm{p}$ parameters (with some exceptions, discussed below).
We apply the same technique to calculate $\chi$.

A nonzero transition dipole ($\chi\ne 0$) in III-V semiconductors modifies the heavy-hole spin-orbit interactions for a two-dimensional hole gas in an asymmetric quantum well.
Here, we study the heavy-hole spin-orbit couplings for a simple (single-subband) valence-band model.
We find a spin-orbit coupling term in the heavy-hole subspace that is linear in the wavevector ($k_{\parallel}$), with Dresselhaus symmetry, and that is proportional to $\chi$.
Because it originates from the finite interband transition dipole matrix element, we call this term dipolar spin-orbit coupling.
From this analysis, we find the dipolar spin-orbit coupling is of the same order as other known contributions.

Including the dipolar spin-orbit coupling in our analysis, we are able to characterize the heavy-hole spin splitting in a triangular GaAs quantum well [the triangular well is defined by the potential give in Eq.~\eqref{eq:triangularpotential}, below].
The dipolar spin-orbit coupling can affect the interpretation of experiments sensitive to the Dresselhaus spin-orbit coupling. 
In recent experiments, the leakage current passing through a double quantum dot containing holes in the Pauli-spin-blockade regime was measured as a function of the direction of an applied magnetic field (angle relative to the current) \cite{wang2016anisotropic}.
The angular dependence of the observed current was consistent with spin-orbit coupling that is almost entirely Dresselhaus-like. 
In other recent experiments, the relaxation time, $T_1$, for a heavy-hole spin in a GaAs quantum dot was measured as a function of the strength of an applied magnetic field, $B$ \cite{bogan2019single}. 
The measured relaxation times are consistent with $T_1 \propto B^{-5}$, which is a signature of Dresselhaus spin-orbit coupling, provided the $k_{\parallel}$-linear Rashba contribution is negligible \cite{bulaev2005spin}.
These two experiments indicate that characterizing the Dresselhaus spin-orbit coupling is crucial to understanding heavy-hole-spin dynamics in these systems. 
Therefore, the dipolar spin-orbit coupling can play an important role in understanding heavy-hole spin dynamics. 

The rest of this article is organized as follows: In Sec.~\ref{sec:kp} we introduce the interband electric dipole coupling into the $\bm{k} \cdot \bm{p}$ formalism going beyond the enevelope-function approximation.
In Sec.~\ref{sec:fpkp} we discuss the first-principles calculation of material parameters (e.~g.~$\bm{k} \cdot \bm{p}$ parameters) for GaAs.
In Sec.~\ref{sec:tripot} we derive the spin-orbit couplings for heavy holes in a triangular quantum well.
Conclusions are given in Sec.~\ref{sec:conclusion}.

\section{Electric-dipole coupling within the $\bm{k} \cdot \bm{p}$ formalism}\label{sec:kp}
The goal of this section is to rederive the $\bm{k}\cdot \bm{p}$ Hamiltonian, extending beyond the envelope-function approximation. 
In this extension, in addition to intraband terms generated within the usual envelope-function approximation, we also include interband coupling generated by the confining potential.

We begin from the Hamiltonian $H = H_0 + U$, where $H_0$ describes a perfectly periodic crystal (including the spin-orbit coupling) and $U = e \bm{E}\cdot\bm{r}$, where $-e$ is the electron charge and $\bm{E}$ is a uniform electric field. 
The term $U$ models the potential experienced by electrons or holes in an asymmetric quantum well at a heterointerface.

The eigenstates of $H_0$ are Bloch waves.
These states can be represented by spinors, $\psi_{\nu\bm{k}}(\bm{r}) = \left[\psi^{\uparrow}_{\nu\bm{k}}(\bm{r}),\psi^{\downarrow}_{\nu\bm{k}}(\bm{r})\right]^{T}$, with components
\begin{equation}
\psi_{\nu\bm{k}}^\sigma(\bm{r}) = \frac{1}{\sqrt{N}}e^{i\bm{k}\cdot\bm{r}}u_{\nu\bm{k}}^\sigma(\bm{r}),
\end{equation}
where $\nu$ is a band index, $\bm{k}$ is a wavevector restricted to the first Brillouin zone, $\sigma$ is a spin index, $N$ is the number of unit cells in the crystal and $u_{\nu\bm{k}}^\sigma(\bm{r})$ are lattice-periodic Bloch amplitudes, which are normalized over the primitive-cell volume, $\Omega$:
\begin{equation}\label{eq:normalization}
\sum_\sigma\int_{\Omega}d^3 r \left|u_{\nu\bm{k}}^\sigma(\bm{r})\right|^2=1.
\end{equation}
The eigenenergies associated with the states $\psi_{\nu\bm{k}}(\bm{r})$ are labeled $\epsilon_{\nu\bm{k}}$.
A convenient complete orthonormal basis can be written in terms of the $\bm{k} = \bm{0}$ Bloch amplitudes, and is given by a set of spinors $\phi_{\nu\bm{k}}(\bm{r}) = \left[\phi^{\uparrow}_{\nu\bm{k}}(\bm{r}),\phi^{\downarrow}_{\nu\bm{k}}(\bm{r})\right]^{T}$, with components \cite{winkler2003spin, luttinger1955motion}, 
\begin{equation}
\phi^{\sigma}_{\nu\bm{k}}(\bm{r}) = \braket{\bm{r} \sigma}{\nu \bm{k}} = \frac{1}{\sqrt{N}}e^{i\bm{k}\cdot\bm{r}}u_{\nu}^\sigma(\bm{r}),
\end{equation}
where $u_{\nu}^\sigma(\bm{r}) := u_{\nu \bm{0}}^\sigma(\bm{r})$. 

Any eigenstate $\ket{\Psi}$ of the Hamiltonian $H$ can be expressed in terms of the states $\ket{\nu' \bm{k}'}$ as 
\begin{equation}\label{eq:eigenstateH}
\ket{\Psi} = \sum_{\nu' \bm{k}'} \Psi_{\nu'}(\bm{k}') \ket{\nu' \bm{k}'}, 
\end{equation}
where $\Psi_{\nu'}(\bm{k}')$ are coefficients that describe the envelope functions in $\bm{k}$-space for a slowly-varying potential.

We insert Eq.~\eqref{eq:eigenstateH} into the Schr\"odinger equation to obtain: 
\begin{equation}\label{eq:SEenv}
\sum_{\nu' \bm{k}'} \Psi_{\nu'}(\bm{k}') \bra{\nu \bm{k}}(H_0 + U)\ket{\nu' \bm{k}'} = \epsilon \Psi_{\nu}(\bm{k}),
\end{equation}
where $\epsilon$ is the eigenenergy of $\ket{\Psi}$. 
We now evaluate the matrix elements $\bra{\nu \bm{k}}(H_0 + U)\ket{\nu' \bm{k}'}$ from Eq.~\eqref{eq:SEenv} to obtain a matrix equation for the envelope functions. 
Within the double-group formulation of $\bm{k} \cdot \bm{p}$ theory \cite{elder2011double}, we write
\begin{equation}\label{eq:matelemH0}
\bra{\nu \bm{k}}H_0 \ket{\nu' \bm{k}'} = \delta_{\bm{k}\bm{k}'} \left[ \left( \epsilon_{\nu\bm{0}} + \frac{\hbar^2 k^2}{2m}\right) \delta_{\nu \nu'} + \frac{\hbar}{m}\bm{k} \cdot \bm{\pi}_{\nu \nu'}\right], 
\end{equation}
with
\begin{equation}\label{eq:pi}
\bm{\pi}_{\nu \nu'} = \sum_{\sigma}\int_{\Omega} d\bm{r} u_{\nu}^{\sigma \ast}(\bm{r})\left[\bm{p} + \frac{\hbar}{2 m c^2} \bm{S} \times \nabla V_0(\bm{r}) \right] u^{\sigma}_{\nu' }(\bm{r}),
\end{equation}
where $V_0$ is the periodic crystal potential and $\bm{S}$ is the electron spin-1/2 operator.
The term $\frac{\hbar}{m}\bm{k}\cdot\bm{\pi}_{\nu\nu'}$ in Eq.~\eqref{eq:matelemH0} is taken as a perturbation in $\bm{k} \cdot \bm{p}$ theory.
In contrast to the single-group formulation of $\bm{k} \cdot \bm{p}$ theory, where the spin-orbit coupling is also a perturbation \cite{winkler2003spin},  here we include the spin-orbit coupling in the unperturbed portion of the $\bm{k} \cdot \bm{p}$ Hamiltonian, so that $\epsilon_{\nu\bm{0}}$ is defined by \cite{elder2011double}
\begin{equation}\label{eq:doublegroup}
    \left[\frac{p^2}{2m} + V_0 + \frac{\hbar}{2m^2c^2} \bm{p} \cdot \bm{S} \times (\nabla V_0)\right]\ket{\nu \bm{0}} = \epsilon_{\nu\bm{0}}\ket{\nu \bm{0}}.
    \end{equation}

We write the matrix elements of $U$ using standard manipulations as
\begin{multline}
\bra{\nu \bm{k}} U \ket{\nu' \bm{k}'} \simeq\frac{e}{V}\delta_{\nu \nu'}\int d\bm{r} e^{i(\bm{k}'-\bm{k})\cdot \bm{r}}\bm{E}\cdot\bm{r}\\
+\frac{e}{N}\bm{E} \cdot\sum_{\bm{K} \sigma}\delta_{\bm{k}-\bm{k}',\bm{K}}  \int_{\Omega} d\bm{r} e^{i\bm{K}\cdot \bm{r}} u_{\nu}^{\sigma\ast}(\bm{r}) \bm{r}u_{\nu'}^{\sigma}(\bm{r}),
\end{multline}
where $V=N\Omega$ is the crystal volume and where the first term gives the usual envelope-function approximation \cite{winkler2003spin}. 
This term results after assuming that the slowly varying envelope functions have substantial Fourier components only for $k,k'\ll \pi/a$.  
In the same limit, only the $\bm{K}=0$ contribution to the second term is relevant, giving
\begin{eqnarray}
\bra{\nu \bm{k}} U \ket{\nu' \bm{k}'} & \simeq & \frac{e}{V} \delta_{\nu \nu'} \int d\bm{r} e^{i(\bm{k}'-\bm{k})\cdot \bm{r}} \bm{E}\cdot\bm{r}  \nonumber \\ 
& + &  ea_B \bm{E} \cdot \bm{d}_{\nu \nu'} \delta_{\bm{k}\bm{k}'}  \label{eq:matelemU3},
\end{eqnarray} 
where $-ea_B\bm{d}_{\nu\nu'}$ is the dipole matrix element, with
\begin{equation}\label{eq:d}
\bm{d}_{\nu \nu'}=  \sum_{\sigma}\int_{\Omega} d\bm{r} u_{\nu}^{\sigma\ast}(\bm{r})\bm{r}u_{\nu'}^{\sigma}(\bm{r})/a_B,
\end{equation}
and $e a_B \simeq 2.5 \, \mathrm{D}$ ($a_B$ is the Bohr radius and D is a debye).  We note that both terms in Eq.~\eqref{eq:matelemU3} arise at leading (zeroth) order in the same small parameter, $|\bm{k}-\bm{k}'|a\ll 1$.

To derive an effective Schr\"odinger equation for the envelope functions, we insert Eqs.~\eqref{eq:matelemH0} and \eqref{eq:matelemU3} into Eq.~\eqref{eq:SEenv}. 
We then write the envelope functions in position space using
\begin{equation}
\Psi_{\nu}(\bm{k}) = \frac{1}{\sqrt{V}}\int d\bm{r} e^{-i \bm{k} \cdot \bm{r}} \Psi_{\nu}(\bm{r}).
\end{equation}
The resulting Schr\"odinger equation is 
\begin{equation}\label{eq:kpHamenv}
\sum_{\nu'}\left[ H^{\nu \nu'}_{\bm{k}\cdot \bm{p}}  + H^{\nu \nu'}_{E} \right]\Psi_{\nu'}(\bm{r}) 
 = \epsilon \Psi_{\nu}(\bm{r}),
\end{equation}
where 
\begin{equation}\label{eq:kpHam}
H^{\nu\nu'}_{\bm{k}\cdot \bm{p}} = \left[\epsilon_{\nu\bm{0}} + \frac{\hbar^2 k^2}{2m} + e\bm{E}\cdot\bm{r} \right] \delta_{\nu \nu'} + \frac{\hbar}{m} \bm{k} \cdot \bm{\pi}_{\nu \nu'},
\end{equation}
 with $\bm{k} \rightarrow -i\nabla$, and 
\begin{equation}\label{eq:HE}
    H^{\nu \nu'}_{E} = ea_B \bm{E} \cdot \bm{d}_{\nu\nu'}.
\end{equation}
Eq.~\eqref{eq:kpHamenv} acts as a Schr\"odinger equation for the long-range degrees of freedom, described by the envelope functions, $\Psi_{\nu}(\bm{r})$.
While the standard envelope-function approximation neglects interband transitions generated by the potential $U$, here they are included via $H_E^{\nu\nu'}$.

In the rest of this paper, we determine the size of the electric-dipole term [Eq.~\eqref{eq:HE}] for the heavy-hole and light-hole bands of GaAs from first principles, and analyze some consequences of this term for an asymmetric quantum well.
The general methods described in this section can, however, be applied to a wide range of other materials and bands, whenever the matrix elements $\bm{d}_{\nu\nu'}$ are nonzero.

\section{First-principles material parameters for G\MakeLowercase{a}A\MakeLowercase{s}}\label{sec:fpkp}

To find general material parameters (e.~g.~$\bm{\pi}_{\nu \nu'}, \bm{d}_{\nu\nu'}$), we need an accurate description of the Bloch amplitudes, $u_{\nu}(\bm{r})$ [see Eqs.~\eqref{eq:pi} and \eqref{eq:d}]. 
These parameters have been calculated for GaAs using various techniques, including tight-binding methods \cite{lewyanvoon1993tight} and density functional theory (DFT) with empirical pseudopotentials \cite{gorczyca1991pseudopotential, pugh1999band} (or a combination of both, where the tight-binding parameters are calculated within DFT \cite{tan2013empirical}).
In pseudopotential methods, the core electrons around each nucleus are `frozen'.
Interactions between valence-shell and core-shell electrons are then included at the level of an effective potential.
In contrast, all-electron DFT techniques solve for the (Kohn-Sham) orbitals of all electrons (including the core electrons). 
An all-electron approach may be important to describe the electronic states at short length scales, close to the nuclei.
All-electron DFT can be performed using, e.g., the open-source {\sc elk} code \cite{elk, elkwebsite}.  The Kohn-Sham orbitals resulting from {\sc elk} have been used to accurately calculate hyperfine parameters for electrons and holes in GaAs and silicon \cite{philippopoulos2020first}. 
In this paper, we use an equivalent procedure to evaluate other GaAs material parameters ($\bm{k}\cdot\bm{p}$ parameters and the matrix elements $\bm{d}_{\nu\nu'}$). 

\begin{figure}
\centering
\includegraphics[width=\columnwidth]{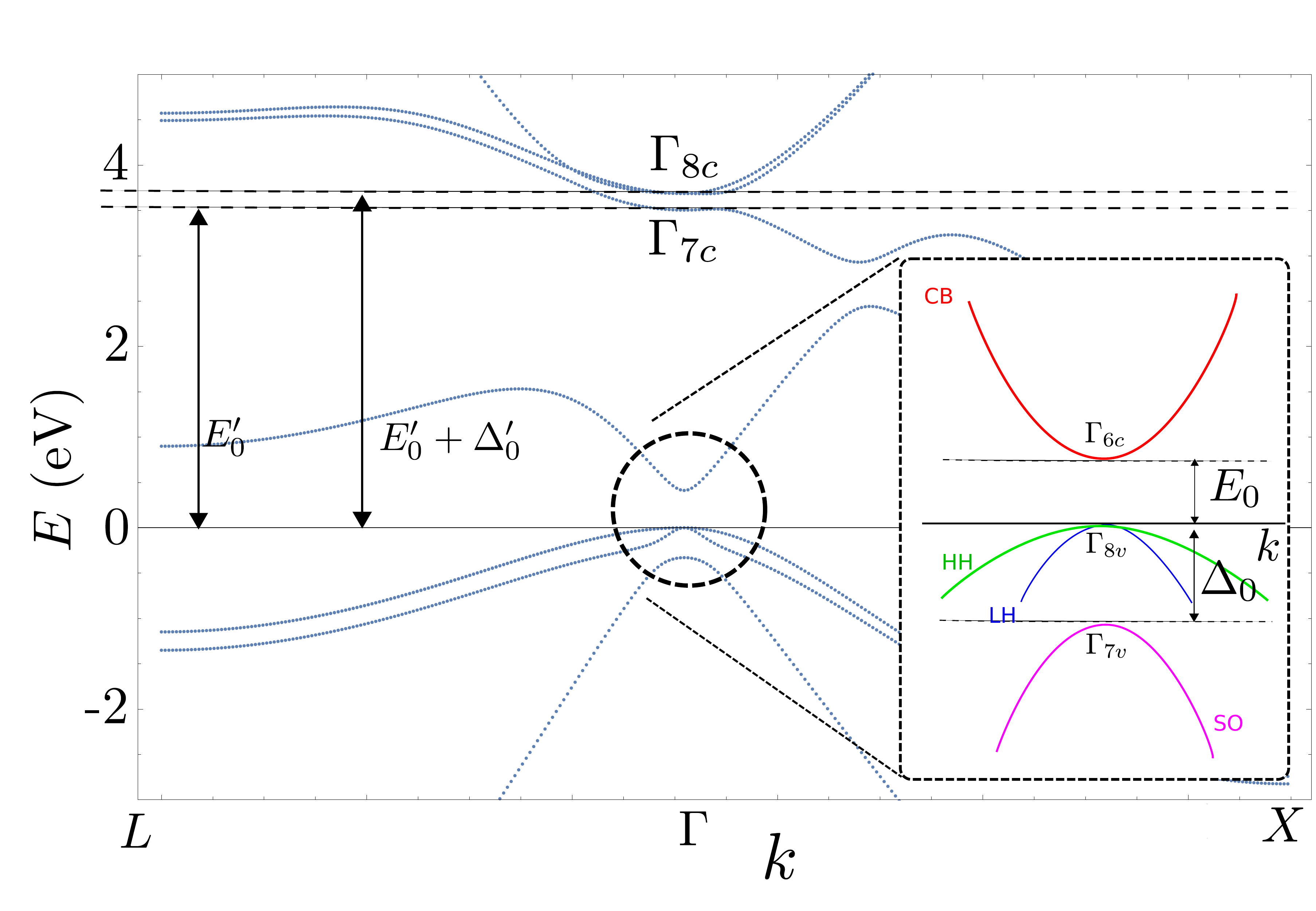}
\caption[Band Structure]{GaAs band structure generated using the {\sc elk} code \cite{elkwebsite,elk}.  The band structure is given between high-symmetry points $L \rightarrow \Gamma \rightarrow X$. The labels $\Gamma_{j_1 j_2}$, $j_1 \in \{6, 7, 8\}$, $j_2 \in \{c, v\}$ for each band at the $\Gamma$ point indicate the representation of the basis states with  the subscript $j_2 = c$ ($v$) indicating conduction- (valence-) band states.  See Sec.~3.3 of Ref.~\cite{winkler2003spin} for more details.
Inset: schematic showing detail of the topmost valence bands and lowest conduction band.
The labels CB, HH, LH, SO indicate the conduction band, heavy-hole band, light-hole band, and split-off band, respectively. }\label{fig:bandstructure}
\end{figure}

From {\sc elk}, we have extracted optimized Kohn-Sham orbitals for GaAs  at the $\Gamma$ point ($\bm{k} = \bm{0}$).
From these orbitals, we have evaluated matrix elements of the momentum operator, giving $\bm{k} \cdot \bm{p}$ parameters $P$, $P'$, and $Q$ (see Appendix \ref{app:conv} for details).
The calculated band structure is shown in Fig.~\ref{fig:bandstructure}, resulting in a first-principles estimate for the band gaps.
In Table \ref{tab:kp}, we compare the $\bm{k} \cdot \bm{p}$ parameters and energy gaps found from this procedure with accepted values (tabulated in the book by Winkler, Ref.~\cite{winkler2003spin}).  We comment on agreement/disagreement of these parameters with the accepted values in Sec.~\ref{sec:Accuracy}, below.

In addition to the $\bm{k}\cdot\bm{p}$ parameters, there are two parameters $(\kappa,q)$ that characterize the strength and symmetry of the Zeeman interaction. 
These parameters are derived as follows.  
In the presence of a magnetic field $\bm{B}$, the electron Zeeman Hamiltonian is
\begin{equation}
H_Z = - \boldsymbol{\mu} \cdot \bm{B},
\end{equation}
where $ \boldsymbol{\mu}$ is the magnetic moment,
\begin{equation}\label{eq:magmom}
 \boldsymbol{\mu} = -\frac{\mu_B}{\hbar}\left(g_L \bm{L} + g_s \bm{S}\right).
\end{equation}
In Eq.~\eqref{eq:magmom}, $\mu_B$ is the Bohr magneton, $g_L = 1$ and $g_s = 2$ are g-factors, $\bm{L}$ is the electron orbital angular momentum operator and $\bm{S}$ is the electron spin operator. 
In the valence band of GaAs, where $\nu$ and $\nu'$ span the heavy-hole and light-hole states ($\Gamma_8$ representation of the $T_d$ double group), the effective Zeeman Hamiltonian matrix is \cite{winkler2003spin, luttinger1956quantum}:
\begin{equation}
{\cal{H}}_Z = -2\kappa\mu_B \bm{B}\cdot\mathbf{J} - 2q\mu_B\bm{B}\cdot{\cal{J}},
\end{equation}
where $\mathbf{J}=(J_x,J_y,J_z)$ is the vector of spin-$3/2$ matrices, ${\cal{J}}=(J_x^3,J_y^3,J_z^3)$, and the parameters $\kappa$ and $q$ are  determined through matrix elements of the magnetic moment operator $\bm{\mu}$.
The parameters $\kappa$ and $q$, calculated from first principles, are compared with the accepted values in Table~\ref{tab:kp}. 

\begin{table}
 \begin{ruledtabular}
 \begin{center}
\begin{tabular}{ c c c }
  Parameter & Present Work & Winkler (Ref.~\cite{winkler2003spin}) \\ 
  \hline
 $P$ & $7.9 \, \mathrm{eV} \angstrom$ & $10.493 \, \mathrm{eV} \angstrom$ \\
 $Q$ & $7.4\, \mathrm{eV} \angstrom$ & $ 8.165 \, \mathrm{eV} \angstrom$ \\  
 $P'$ & $1.7i \, \mathrm{eV} \angstrom$ & $4.780i \, \mathrm{eV} \angstrom$ \\
 $E_0$ & $0.41\, \mathrm{eV}$  & $1.519\, \mathrm{eV}$ \\
 $\Delta_0$ & $0.33\, \mathrm{eV}$ & $0.341 \, \mathrm{eV}$ \\
 $E_0'$ & $3.5\, \mathrm{eV} $ & $4.488 \, \mathrm{eV} $ \\
 $\Delta_0'$ & $0.18\, \mathrm{eV} $ & $0.171 \, \mathrm{eV} $ \\
 $\kappa$ & $-0.61 $ & $1.20$ \\
 $q$ & $-0.01 $ & $0.01$ \\
 $\chi$ & 0.2 & -
\end{tabular}
\caption{Material parameters found from density functional theory (present work) and the accepted values from Table D.1 of Ref.~\cite{winkler2003spin}. The $\bm{k} \cdot \bm{p}$ parameters $P$, $Q$, and $P'$ are defined in Eq.~(3.3) of Ref.~\cite{winkler2003spin}. For a discussion of the convergence of the calculation, see Appendix~\ref{app:conv}.}\label{tab:kp}
\end{center}
\end{ruledtabular}
\end{table}

As described above, matrix elements of the momentum operator ($\propto P,P',Q$), band gaps ($E_0,E_0',\Delta_0,\Delta_0'$), and matrix elements of the magnetic moment operator ($\propto \kappa,q$) are sufficient to parametrize $\bm{k}\cdot\bm{p}$ theory within the envelope-function approximation.  
However, as described in Sec.~\ref{sec:kp}, in certain circumstances, matrix elements of the electric-dipole operator ($\propto \bm{d}_{\nu\nu'}$) may also be relevant. In a subspace spanned by states that transform according to the $\Gamma_8$ representation of the tetrahedral double group (e.g., the heavy-hole and light-hole states), the effective projected electric-dipole Hamiltonian matrix derived from $H_E^{\nu\nu'}$  is \cite{bir1963spin}:
\begin{equation}\label{eq:symdip}
{\cal{H}}_E = \frac{1}{\sqrt{3}}ea_B\chi\left[E_x \{J_y,J_z\} + E_y \{J_z,J_x\} +E_z \{J_x,J_y\}\right],
\end{equation}
where $\chi$ is a parameter that controls the strength of the electric-dipole matrix elements.
We have numerically evaluated the matrix elements $\bm{d}_{\nu\nu'}$ [see Eq.~\eqref{eq:HE}] giving $\chi=0.2$ (listed in Table \ref{tab:kp}).
While ${\cal H}_E$ arises physically from the electric-dipole operator, this term can be written as an effective pseudospin $J=3/2$ quadrupole term (similar to that analyzed in Refs.~\cite{winkler2004spin, culcer2006spin}).\footnote{Specifically, we can rewrite Eq.~\eqref{eq:symdip} as ${\cal H}_E =  e a_B\chi \left(E_x \tilde{Q}_{yz} + \mathrm{c.p.}\right)$. Here, ``$\mathrm{c.p.}$'' indicates cyclic permutations and $\tilde{Q}_{ij}=\{J_i,J_j\}$ is proportional to the usual quadrupole matrix up to an additive constant \cite{abragam1961principlesp163}: $Q_{ij}=eQ\left[\frac{3}{2}\tilde{Q}_{ij}-\delta_{ij}J(J+1)\right]/\left[6J(2J-1)\right]$, with quadrupole moment $Q$.  Thus, although this term is generated through an electric dipole coupling, the usual dipole selection rules do not apply (with respect to the pseudospin).  In particular, $\mathcal{H}_E$ allows for ``double-quantum'' transitions having $|\Delta J_z|=2$ (e.g. $\ket{J_z=+3/2}\leftrightarrow\ket{J_z=-1/2}$). }

The electric-dipole term given by Eq.~\eqref{eq:symdip} may lead to important measurable effects. 
For example, when this term is included, an oscillating electric field can drive electric-dipole transitions between heavy-hole and light-hole states.
For crystals with a center of inversion symmetry, $\chi=0$ identically and these transitions vanish.
However, in GaAs we find (within DFT) that the heavy-hole/light-hole transition dipole is $ea_B\chi \simeq 0.5\, \mathrm{D}$, a substantial fraction ($\simeq 40\%$) of that for a $1s\to 2p$ hydrogen atom transition.\footnote{
The electric-dipole matrix element describing this hydrogenic transition is given by $e\bra{1\,0\,0}x\ket{2\,1\,\pm1}=p\chi_{1s\rightarrow2p}$, where $\chi_{1s\rightarrow2p} \approx 0.5$  ($\chi/\chi_{1s\rightarrow2p} \approx 0.4$) and the states, $\ket{n\,l\,m}$ are the hydrogen-atom eigenstates ($n$ is the principal quantum number, $l$ is the orbital angular momentum, and $m$ is the orbital angular momentum along the axis of quantization).
}
In Sec.~\ref{sec:tripot} below, we explore further consequences of this term for an asymmetric quantum well.

\subsection{Accuracy of first-principles parameters}
\label{sec:Accuracy}

The results given in Table \ref{tab:kp} show a broad range of agreement between the parameters calculated here within DFT and the accepted values listed in Ref.~\cite{winkler2003spin}.  
The spin-orbit gaps, $\Delta_0$ and $\Delta_0^\prime$, are well-reproduced in DFT (within 5\% of the accepted values).  
However, $P, Q$, and $E_0'$ deviate by 20\%-30\%, and the calculated $P'$ and $E_0$ differ from the accepted values by as much as a factor of 4. 
We note that the spin-orbit gaps $\Delta_0,\Delta_0'$ depend on short-range properties of the wavefunctions near the nuclear cores (where the spin-orbit coupling diverges $\propto \nabla V(r)\propto 1/r^2$).  
This level of agreement is consistent with the agreement found previously for short-range hyperfine parameters \cite{philippopoulos2020first}.  
In contrast, the other parameters listed above depend on the electronic structure far from the nuclear cores.  For example, in a tight-binding theory, the band gaps $E_0$ and $E_0'$ depend on overlaps of atomic wavefunctions localized at different sites.  These band gaps are known to be underestimated within DFT under the local density approximation \cite{perdew1985density}. 
In addition, while we find that the calculated values for $\kappa$ and $q$ deviate from the accepted values, the calculated value for $\kappa$ nearly coincides with the value expected in the tight-binding limit, $\kappa =-2/3$ \cite{luttinger1956quantum}.

The discussion above indicates a strong degree of confidence in the accuracy of the DFT procedure in calculating short-range quantities.  
For longer-range quantities, the results are mixed but nevertheless produce the correct order of magnitude: With the exception of $\kappa$ and $q$, the calculated quantities are all within a factor of $\sim 4$ of their accepted value.
We therefore expect the true value of the parameter $\chi$ to be within a factor of $\sim 4$ of the value reported here.
To unambiguously establish the accuracy of the calculated value of $\chi$, a direct comparison to experiment is required.
As discussed above, observation of electrically driven heavy-hole/light-hole Rabi oscillations would provide a direct measurement of $\chi$.
Alternative possible experiments that could quantitatively establish $\chi$ would be measurements of the heavy-hole spin splitting and spin-orbit coupling, which we discuss in the next section.

\section{Spin-orbit interactions for a triangular quantum well} \label{sec:tripot}

The goal of this section is to explore the influence of ${\cal{H}}_E$ [Eq.~\eqref{eq:symdip}] on heavy-hole spin-orbit coupling for an asymmetric quantum well.
We consider a quantum well formed at a heterointerface where the confinement can be described by a triangular potential due to an electric field $\bm{E}=E_z\hat{\bm{z}}$,
\begin{equation}\label{eq:triangularpotential}
U(z) = \begin{cases} 
  \infty&\quad z\le 0,\\
  e E_z z&\quad z>0.\\
\end{cases}
\end{equation}   
For the valence band of a III-V semiconductor, where $\nu$ and $\nu'$ are restricted to the heavy-hole (HH) and light-hole (LH) states, we write the Hamiltonian matrix (for a positively charged hole with positive effective mass)
\begin{equation}\label{eq:fullH}
{\cal{H}}={\cal{H}}_L - U(z)I_4 + {\cal{H}}_1 + {\cal{H}}_3 - {\cal{H}}_E,
\end{equation}
where $I_4$ is the $4\times 4$ identity matrix and
\begin{equation}\label{eq:LuttingerHam}
{\cal{H}}_L=\frac{\hbar^2}{2m}\left[\left(\gamma_1+\frac{5}{2}\gamma_2\right)k^2 I_4-2\gamma_2(\bm{k}\cdot\mathbf{J})^2\right]
\end{equation}
is the Luttinger Hamiltonian within the spherical approximation, with parameters $\gamma_1$ and $\gamma_2$ (see Table \ref{tab:params} for values in GaAs). 
The term ${\cal{H}}_1$ is linear in $k$ \cite{cardona1988relativistic}:
\begin{equation}\label{eq:H1}
{\cal{H}}_1 = - \frac{2 C_k}{\sqrt{3}} \left[ k_x \{ J_x, J_y^2-J_z^2 \} + \mathrm{c.p.}\right],
\end{equation}
 while ${\cal{H}}_3$ is cubic in $k$:
\begin{eqnarray}\label{eq:H3}
{\cal{H}}_3 &=& - b_{41}\left[\{k_x, k_y^2 - k_z^2\} J_x + \mathrm{c.p.}\right] \nonumber\\ 
	& - & b_{42} \left[\{k_x, k_y^2 - k_z^2\} J^3_x + \mathrm{c.p.}\right]\nonumber \\
	& - & b_{51} \left[\{k_x, k_y^2 + k_z^2\} \{J_x, J_y^2 - J_z^2\} + \mathrm{c.p.}\right]\nonumber \\
	& - & b_{52} \left[k_x^3 \{J_x, J_y^2 - J_z^2\} + \mathrm{c.p.}\right],
\end{eqnarray}
where $C_k$, $b_{41}$, $b_{42}$, $b_{51}$, and $b_{52}$ are material parameters (see Table \ref{tab:params} for values in GaAs).
In Eqs.~\eqref{eq:LuttingerHam}, \eqref{eq:H1}, and \eqref{eq:H3}, $\bm{k}$ is a differential operator, $\bm{k} = -i\nabla$.
In contrast to ${\cal{H}}_1$ and ${\cal{H}}_3$, ${\cal{H}}_E$ depends on the strength of the electric field, $E_z$ [see Eq.~\eqref{eq:symdip}].
However, ${\cal{H}}_E$, ${\cal{H}}_1$, and ${\cal{H}}_3$ share a common origin: they all stem from bulk-inversion asymmetry.
Therefore, a general theory accounting for bulk-inversion asymmetry should include ${\cal{H}}_E$.

\begin{table}
 \begin{ruledtabular}
 \begin{center}
\begin{tabular}{ c c c c c c c c}
 $\gamma_1$ & $\gamma_2$ & $C_k$ & $b_{41}$ & $b_{42}$ & $b_{51}$ & $b_{52}$ & $r_{41}$ \\
  &  & (eV$\angstrom$) & (eV$\angstrom^3$) & (eV$\angstrom^3$) & (eV$\angstrom^3$) & (eV$\angstrom^3$) & (e$\angstrom^2$)\\
 \hline
  $6.85$ & $2.10$ & $-0.0034$ & $-81.93$ & $1.47$ & $0.49$ & $-0.98$ & $-14.62$
\end{tabular}
\caption{GaAs valence-band parameters.
The parameters $\gamma_1$ and $\gamma_2$ are taken from Table D.1 of Ref.~\cite{winkler2003spin}; $C_k$, $b_{41}$, $b_{42}$, $b_{51}$, and $b_{52}$ are taken from Table 6.3 of Ref.~\cite{winkler2003spin}. The parameter $r_{41}$ is taken from Table 6.6 of Ref.~\cite{winkler2003spin}. }\label{tab:params}
\end{center}
\end{ruledtabular}
\end{table}

To derive the effective heavy-hole spin-orbit Hamiltonian, we rewrite ${\cal{H}}$ as 
\begin{equation}\label{eq:ptHam}
    {\cal{H}} = {\cal{H}}_0 + {\cal{H}}',
\end{equation}
where ${\cal{H}}_0$ contains the potential energy, $U$, and the diagonal part of the Luttinger Hamiltonian, ${\cal{H}}_L$, giving matrix elements
\begin{equation}
    {\cal{H}}^{\nu\nu'}_0 = \left[({\cal{H}}_L)_{\nu\nu} - U(z)\right]\delta_{\nu\nu'}.
\end{equation}
The eigenfunctions of ${\cal{H}}_0^{\nu\nu}$ are  $\Psi^{k_x,k_y,n}_{\nu}(\bm{r}) = F^{k_x}(x)F^{k_y}(y)F^n_{\nu}(z)$, where
$F^{k_{x}}(x)$ and $F^{k_{y}}(y)$ are plane waves, while the envelope function $F_{\nu}^n(z)$ solves the differential equation:
\begin{equation}\label{eq:SE}
\left[-\frac{\hbar^2}{2m_{\nu}}\frac{d^2}{dz^2} - U(z)\right]F_{\nu}^n(z) = \epsilon_{\nu}^n F_{\nu}^n(z).
\end{equation}
Here, $m_{\nu}$ is the effective mass ($m_{\mathrm{HH}}=\frac{m}{\gamma_1-2\gamma_2}$ for heavy holes and $m_{\mathrm{LH}}=\frac{m}{\gamma_1+2\gamma_2}$ for light holes), and $\epsilon_{\nu}^n$ is the energy for subband $n$. 
The envelopes,  $F_{\nu}^n(z)$, are given by Airy functions (see Appendix \ref{app:solutionstri}).

We project the Hamiltonian matrix ${\cal{H}}$ onto the lowest subband ($n=1$) to obtain a $4 \times 4$ Hamiltonian matrix, $\hat{H}$ (see Appendix \ref{app:higherbands} for a discussion of the influence of higher subbands, $n>1$).
The `hat' on $\hat{H}$ indicates that the matrix describes only the lowest subband, $n=1$.
The matrix elements of $\hat{H}$ can be written in terms of diagonal matrix elements of $k_z^2$, e.g.,
\begin{equation}\label{eq:kz2}
\left<k_z^2\right> = -\int_0^{\infty} dzF^1_{\mathrm{HH}}(z)\frac{d^2}{dz^2}F^1_{\mathrm{HH}}(z),
\end{equation}
together with the following parameters:
\begin{equation}\label{eq:l}
l = -\frac{\sqrt{3}\hbar^2}{2m}\xi \gamma_2
\end{equation}
and
\begin{equation}\label{eq:lambda}
\lambda = \frac{\sqrt{3}\hbar^2}{m}\gamma_2\eta_1,
\end{equation}
where
\begin{equation}\label{eq:xi}
\xi = \int_0^{\infty} dzF^1_{\mathrm{HH}}(z)F^1_{\mathrm{LH}}(z),
\end{equation}
and 
\begin{equation}\label{eq:eta}
\eta_i = \int_0^{\infty} dzF^1_{\mathrm{HH}}(z)\frac{d^i}{dz^i}F^1_{\mathrm{LH}}(z),
\end{equation}
where $\frac{d^i}{dz^i}$ is the $i^{\mathrm{th}}$ derivative with respect to $z$.

Using second-order degenerate perturbation theory (an approximate Schrieffer-Wolff transformation \cite{schrieffer1966relation}), we project the Hamiltonian matrix, $\hat{H}$, onto the heavy-hole subspace.
This gives the effective $2 \times 2$ heavy-hole Hamiltonian: 
\begin{equation}\label{eq:2x2Hamiltonian}
\hat{H}_{\mathrm{HH}} = \epsilon(k_{\parallel}) + \hat{H}_c + \hat{H}_D,
\end{equation}
where $\bm{k}_{\parallel}=k_x\hat{\bm{x}}+k_y\hat{\bm{y}}$, $ \epsilon(k_{\parallel}) = \epsilon^1_{\mathrm{HH}}+(\gamma_1+\gamma_2)\frac{\hbar^2k_{\parallel}^2}{2m}$, $\hat{H}_c$ represents cubic (in $k_{\parallel}$) spin-orbit coupling and $\hat{H}_D$ represents the linear (in $k_{\parallel}$) Dresselhaus spin-orbit coupling.
The model studied here also leads to a linear Rashba spin-orbit coupling (see Appendix~\ref{app:Rashba}).
However, for the range of electric fields considered, we find the linear Rashba spin-orbit coupling to be at most $\sim 10\%$ of the linear Dresselhaus spin-orbit coupling (see Fig.~\ref{fig:rashba} below).
Therefore, for simplicity, we neglect the linear Rashba spin-orbit coupling in the main text and restrict the discussion of this term to Appendix \ref{app:Rashba}.

\subsection{Cubic spin-orbit coupling}\label{sec:cubic}

\begin{figure}
\centering
\includegraphics[width=\columnwidth]{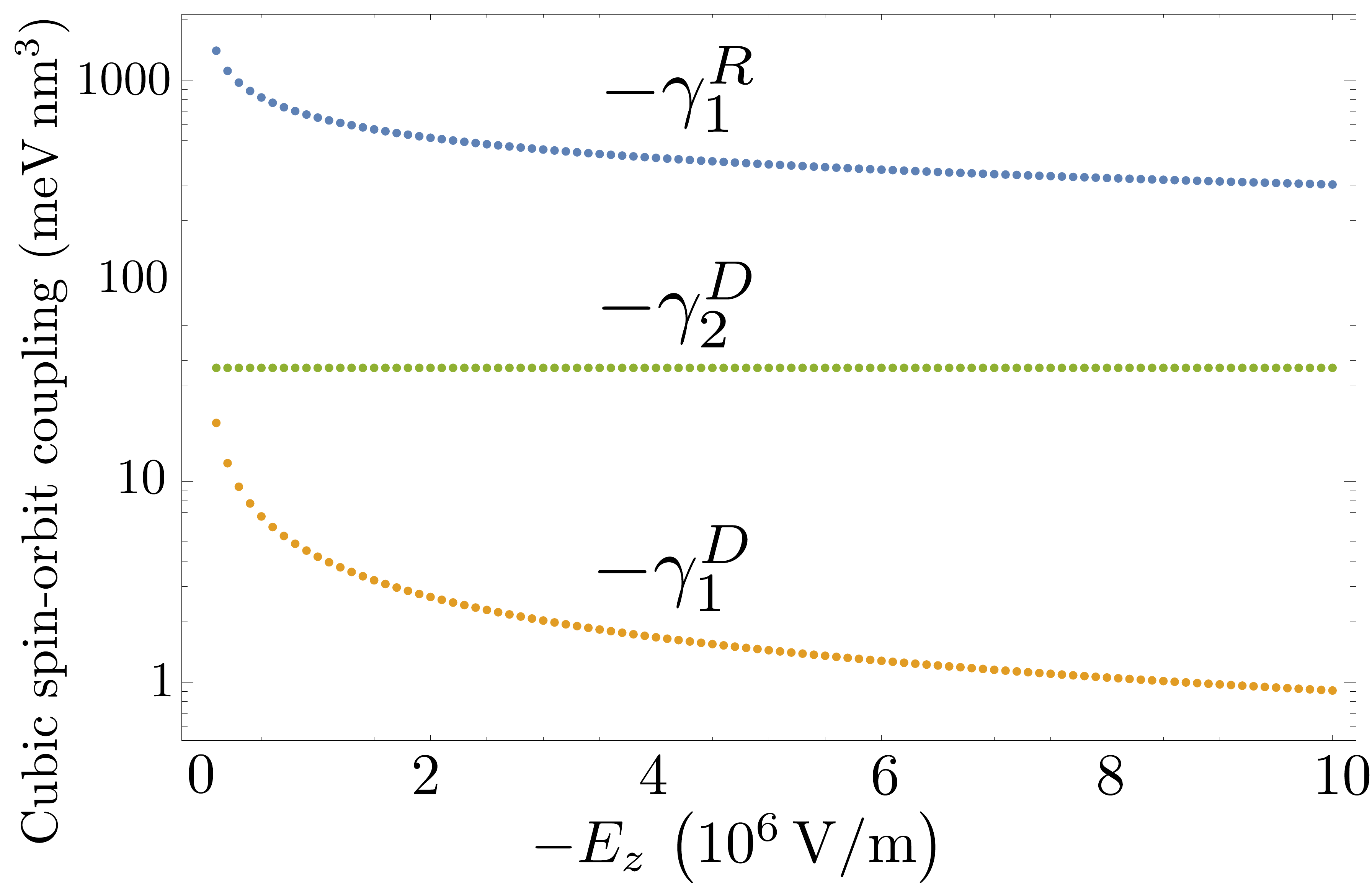}
\caption[Cubic]{Magnitude of the cubic spin-orbit coupling coefficients $\gamma_1^{R}$ [in blue, see Eq.~\eqref{eq:gammaR1}], $\gamma_2^{D}$ [in green, see Eq.~\eqref{eq:gammaD2}], and $\gamma_1^{D}$ [in yellow, see Eq.~\eqref{eq:gammaD1}] as a function of electric field, $E_z$. The material parameters used (for GaAs) are given in Table \ref{tab:params}. }\label{fig:cubic}
\end{figure}

The cubic spin-orbit coupling term is 
\begin{equation}\label{eq:Hc}
\hat{H}_{c} =  i \gamma^R\left(k_+^3\sigma_--k_-^3\sigma_+\right) 
	+ \gamma^D \left( k_+k_-k_+ \sigma_- + k_-k_+k_- \sigma_+ \right),
\end{equation}
where 
$\gamma^R$ and $\gamma^D$ are the cubic Rashba and Dresselhaus spin-orbit couplings, respectively, $k_{\pm} = k_x \pm ik_y$, and $\sigma_{\pm}= (\sigma_{x}\pm i \sigma_y)/2$ are Pauli matrices. 
For the triangular potential chosen here, $\gamma^R \simeq \gamma_1^R$ (see Appendix \ref{app:cubic}), where 
\begin{equation}\label{eq:gammaR1}
\gamma^R_1 = - \frac{2 \lambda l}
			{\Delta_{\mathrm{HL}}}
\end{equation}
and $\gamma^D \simeq \gamma_1^D+\gamma_2^D$, where
\begin{equation}\label{eq:gammaD1}
\gamma_1^D = \frac{l C_k \xi^2}
	{\Delta_{\mathrm{HL}}},
\end{equation}
and
\begin{equation}\label{eq:gammaD2}
\gamma_2^D =  \frac{3 \hbar^2 \gamma_2 }
	{8 m\Delta_{\mathrm{HL}}}\left[\left(4b_{41}+7b_{42}+2b_{51}\right)\xi \eta_2 - 8 b_{51} \eta_1^2\right].
\end{equation}
In Eqs.~\eqref{eq:gammaR1}, \eqref{eq:gammaD1}, and \eqref{eq:gammaD2} $\Delta_{\mathrm{HL}} = \epsilon_{\mathrm{HH}}^1 - \epsilon_{\mathrm{LH}}^1$ is the heavy-hole/light-hole splitting in the lowest subband. 
As can be seen from Fig.~\ref{fig:cubic}, for electric field strengths $\left|E_z\right|\gtrsim 10^6\,\mathrm{V}/\mathrm{m}$, we have $|\gamma^R| \gg |\gamma^D |\simeq |\gamma^D_1+\gamma^D_2|$. 
We thus find that the cubic spin-orbit coupling is predominantly Rashba-like (consistent with, e.g., Ref.~\cite{marcellina2017spin}). 

\subsection{Linear spin-orbit coupling}\label{sec:Dresselhaus}

\begin{figure}
\centering
\includegraphics[width=\columnwidth]{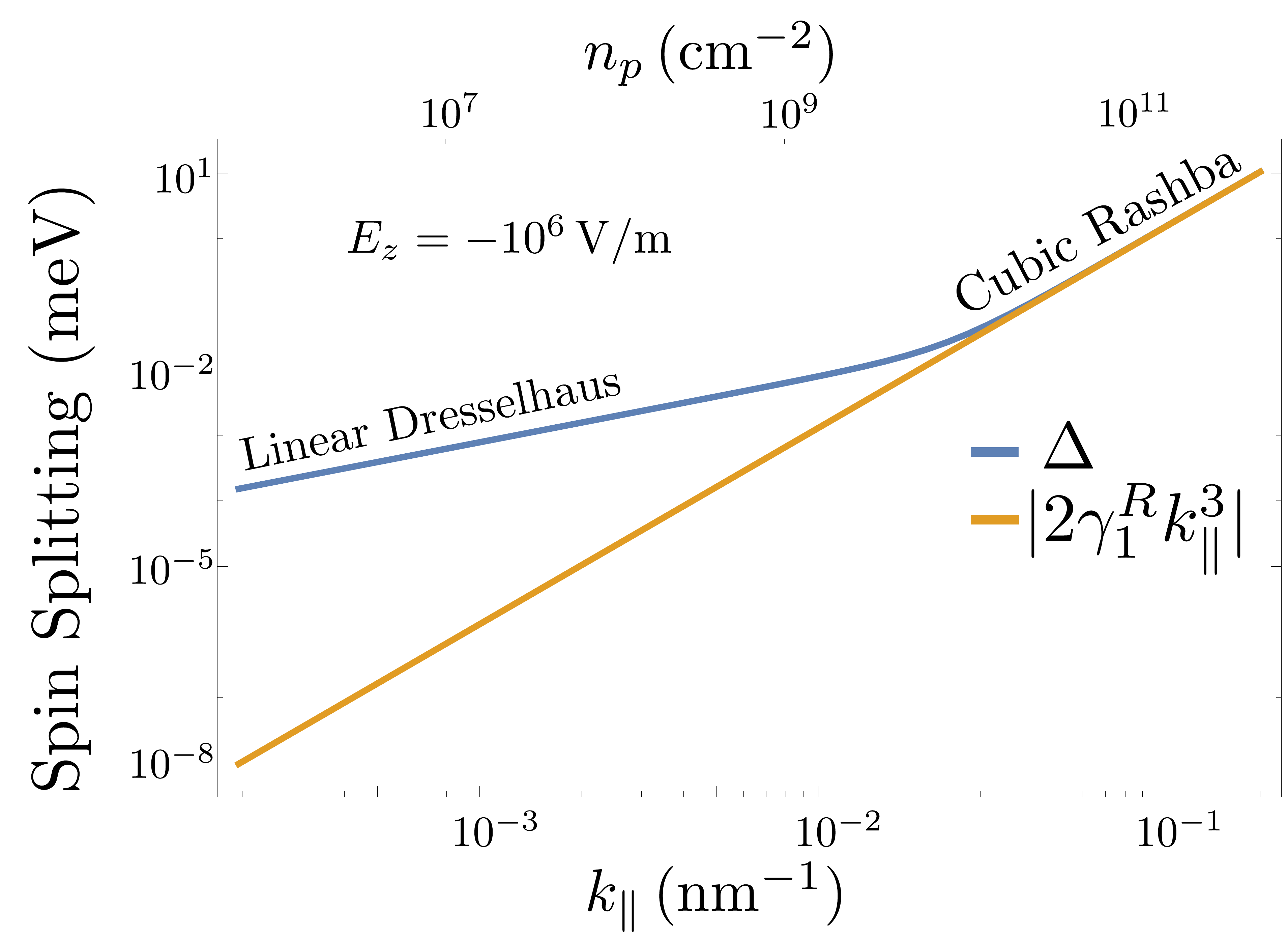}
\caption{Spin splitting, $\Delta$, in GaAs from eigenvalues of Eq.~\eqref{eq:2x2Hamiltonian} (in blue) as a function of $k_{\parallel}=k_x$ ($\bm{k}_{\parallel} = k_{\parallel} \hat{\bm{x}}$) for an electric field $E_z = - 10^6\,\mathrm{V}/\mathrm{m}$.
The top horizontal axis indicates the hole sheet density,  $n_p = k_{\parallel}^2 /(2\pi)$, associated with a Fermi wavevector, $k_{F}=k_{\parallel}$.
The splitting is linear in $k_{\parallel}$ for $k_{\parallel}\lesssim 0.02\,\mathrm{nm}^{-1}$, meaning the linear Dresselhaus spin-orbit coupling dominates in this region. The splitting due to the spin-orbit coupling term with coefficient $\gamma_1^R$ (in yellow) dominates at large $k_{\parallel}$. The material parameters used (for GaAs) are given in Table \ref{tab:params}.}\label{fig:ss}
\end{figure}

Spin-dependent transport and spin-relaxation measurements performed on heavy holes in GaAs quantum dots are consistent with a spin-orbit coupling that is substantially Dresselhaus-like \cite{wang2016anisotropic, bogan2019single}.
As demonstrated in Sec.~\ref{sec:cubic}, when the cubic spin-orbit coupling dominates over the linear spin-orbit coupling (large $k_{\parallel}$ regime), we find a Rashba-like heavy-hole spin-orbit coupling.
However, the linear spin-orbit coupling that dominates for small $k_{\parallel}$ is predominantly Dresselhaus-like [as described above following Eq.~\eqref{eq:2x2Hamiltonian}, see Fig.~\ref{fig:ss}].
This observation may suggest that the experiments conducted in Refs.~\cite{wang2016anisotropic, bogan2019single} are sensitive to the small-$k_\parallel$ spin-orbit coupling analyzed here.

The linear Dresselhaus spin-orbit coupling is 
\begin{equation}\label{eq:HD}
\hat{H}_D = \beta
	\left(k_-\sigma_++k_+\sigma_-\right).
\end{equation}
The coefficient $\beta$ can be written as a sum of five terms:
\begin{equation}\label{eq:beta}
\beta = \beta_{C_k} + \beta_{\chi} + \beta_0 + \beta_{b1} + \beta_{b2},
\end{equation}
where
\begin{equation}\label{eq:betaCk}
\beta_{C_k} = \frac{\sqrt{3}C_k }{2},
\end{equation}
\begin{equation}\label{eq:betachi}
\beta_{\chi} = - \frac{2 \lambda ea_B E_z \chi \xi}
		{\Delta_{\mathrm{HL}}},
\end{equation}
\begin{equation}\label{eq:beta0}
\beta_0 = -\frac{2 \lambda C_k \eta_1}
	{\Delta_{\mathrm{HL}}},
\end{equation}
\begin{equation}\label{eq:betab1}
\beta_{b1} = \frac{3}{4} \left(b_{42} + b_{51} \right) \left<k_z^2\right>,
\end{equation}
and
\begin{equation}\label{eq:betab2}
\beta_{b2} = \frac{\sqrt{3}b_{52}\lambda\eta_3}{\Delta_{\mathrm{HL}}}. 
\end{equation}

\begin{figure} 
\centering
\includegraphics[width=\columnwidth]{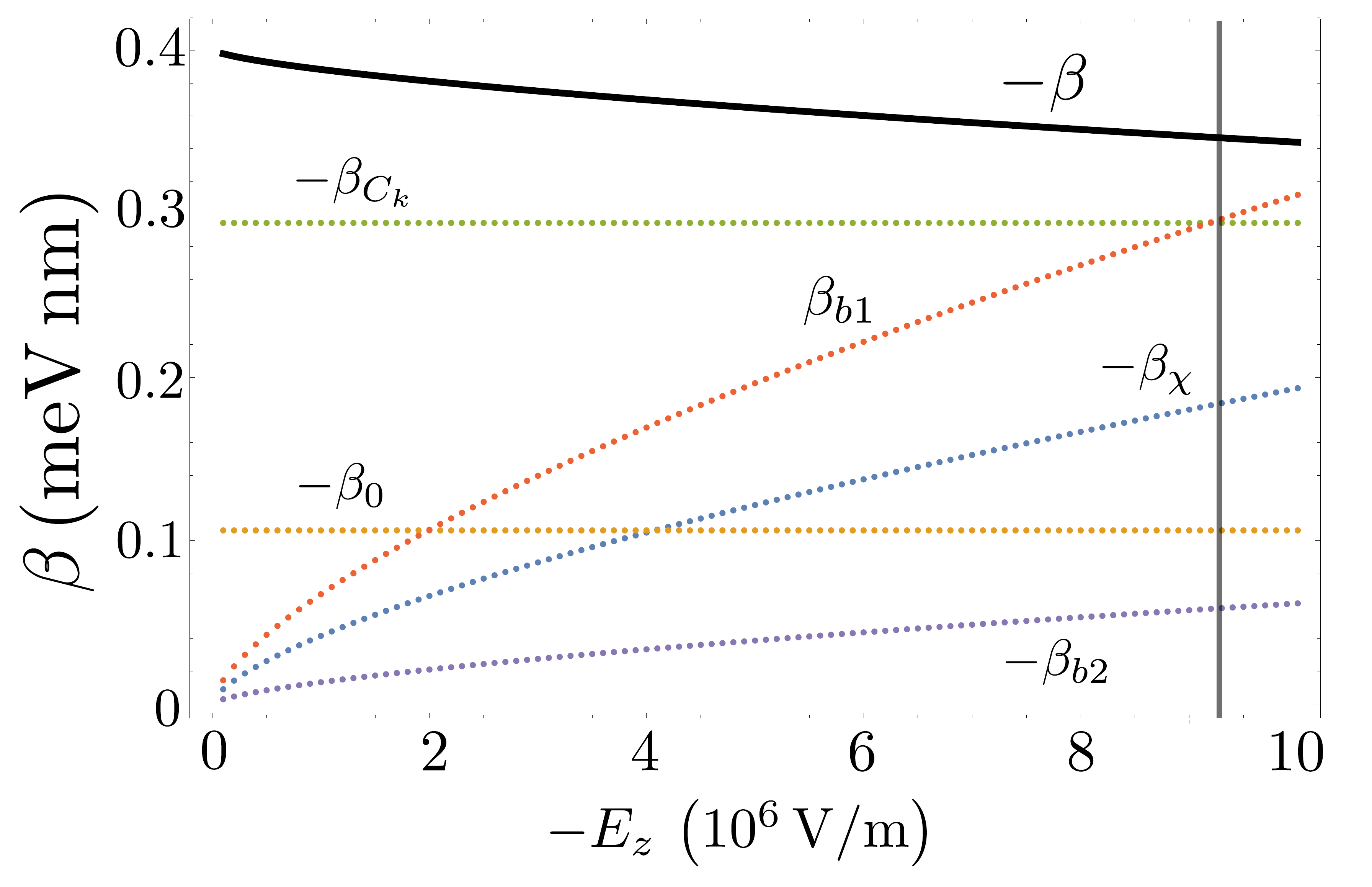}
\caption[dress]{Magnitude of the linear Dresselhaus spin-orbit coupling coefficient, $\beta$ (black solid line) as a function of electric field, $E_z$.  
The dotted lines give the different contributions to $\beta$: $\beta_{C_k}$ [in green, see Eq.~\eqref{eq:betaCk}], $\beta_{b1}$ [in red, see Eq.~\eqref{eq:betab1}], $\beta_{\chi}$ [in blue, see Eq.~\eqref{eq:betachi}], $\beta_{0}$ [in yellow, see Eq.~\eqref{eq:beta0}], and $\beta_{b2}$ [in purple, see Eq.~\eqref{eq:betab2}]. 
 The material parameters used (for GaAs) are listed in Table \ref{tab:params}.
 The gray line indicates the value of $E_z$ for which $\beta_{C_k}+\beta_{b1}=0$.}\label{fig:dress}
\end{figure}

\begin{figure} 
\centering
\includegraphics[width=\columnwidth]{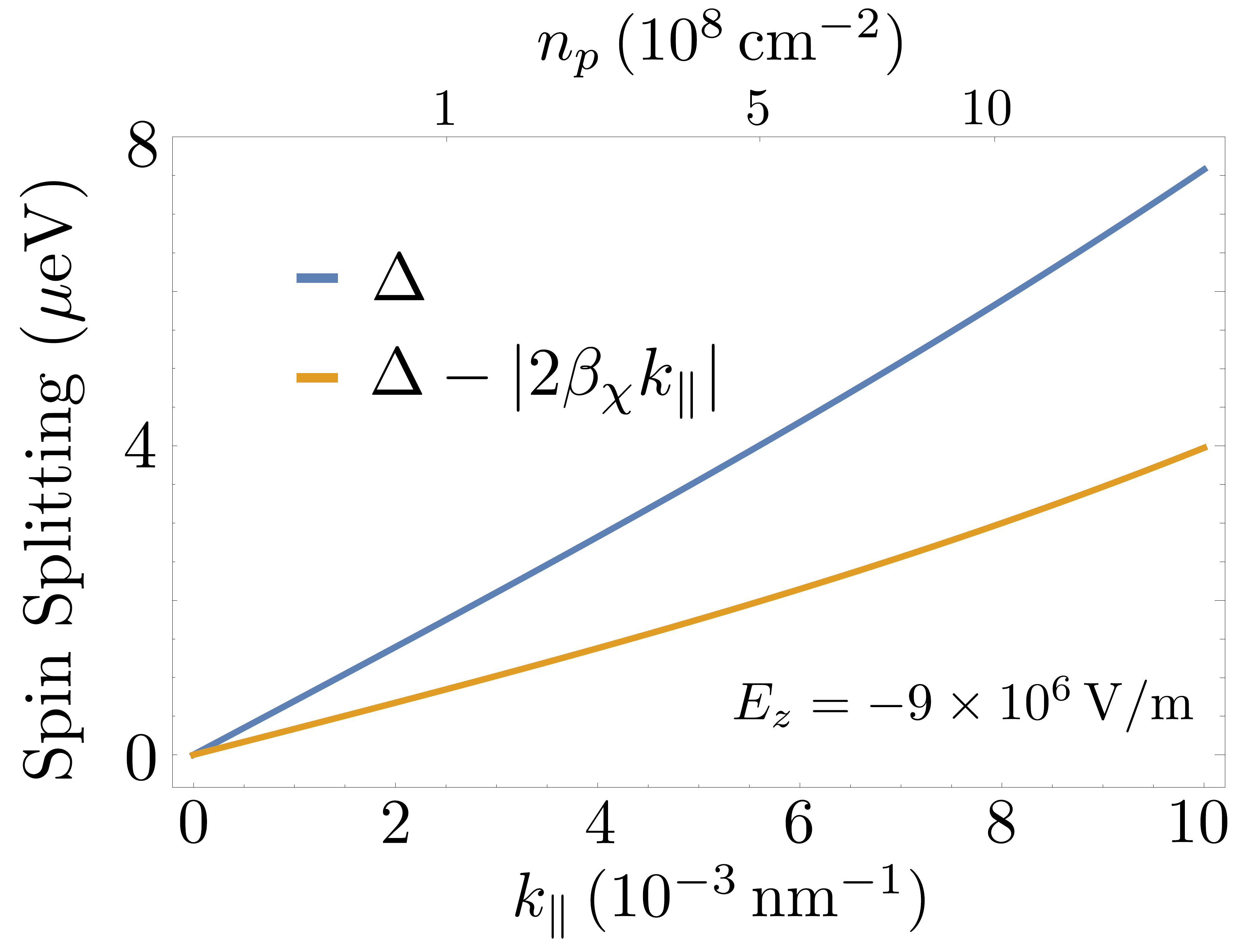}
\caption[dress]
{Total spin splitting, $\Delta$, (blue line) and spin splitting assuming $\chi \to 0$ (yellow line).
For this choice of electric field ($E_z = - 9 \times 10^{6} \, \mathrm{V}/\mathrm{m}$), $\beta_{\chi}$ is the largest contribution to the spin splitting (see Fig.~\ref{fig:dress}). 
The top horizontal axis indicates the hole sheet density,  $n_p = k_{\parallel}^2 /(2\pi)$, associated with a Fermi wavevector, $k_{F}=k_{\parallel}$.
The material parameters used (for GaAs) are displayed in Table \ref{tab:params}.
}\label{fig:ssvanishingchi}
\end{figure}

The dipolar spin-orbit coupling term (the term with coefficient $\beta_{\chi}$) arises from non-vanishing electric-dipole matrix elements (${\cal{H}}_E$ with $\chi \ne 0$).
Therefore, it vanishes identically within the envelope-function approximation.
However, $\beta_\chi$ is of the same order as the other linear Dresselhaus spin-orbit coefficients for $E_z \lesssim - 10^6\,\mathrm{V}/\mathrm{m}$ (see Fig.~\ref{fig:dress}).
As can be seen in Fig.~\ref{fig:dress}, $\beta_{\chi}$, $\beta_{b1}$, and $\beta_{b2}$ vary as a function of electric field. 
In fact, there is a value of the electric field where $\beta_{b1}$ exactly cancels $\beta_{C_k}$ (see gray line in Fig.~\ref{fig:dress}).
For an electric field, $E_z \sim - 9 \times 10^6\, \mathrm{V}/\mathrm{m}$, the dipolar spin-orbit coupling ($\propto \beta_{\chi}$) contributes approximately half of the total spin splitting at small $k_\parallel$, where the linear Dresselhaus term dominates (see Fig.~\ref{fig:ssvanishingchi}). 
The dipolar spin-orbit coupling is therefore necessary for a quantitative theory of spin-orbit couplings for heavy holes in asymmetric GaAs quantum wells. 
Specifically, this term may be relevant in interpreting the experimental results of Refs.~\cite{wang2016anisotropic, bogan2019single}.

The spin states of holes split at finite $k_\parallel$, typically due to \emph{either} structure-inversion asymmetry (e.g., an asymmetric quantum well) \emph{or} due to bulk inversion asymmetry \cite{rashba1988spin}.  The dipolar spin-orbit coupling $\propto \chi E_z$ requires both structure inversion asymmetry (finite $E_z$) and bulk inversion asymmetry (finite $\chi$) in order to contribute to the spin splitting.  The spin splitting has been measured for heavy holes in GaAs quantum wells through Shubnikov-de Haas oscillations \cite{marcellina2018electrical}.
Provided similar measurements could be performed with a typical confining electric field and range of sheet density as given in Fig.~\ref{fig:ssvanishingchi}, the value of $\chi$ determined here could be confirmed.  

\subsubsection{Heavy-hole relaxation time, $T_1$}
In Ref.~\cite{bogan2019single}, the relaxation time $T_1$ has been measured for a quantum-dot-confined heavy-hole spin as a function of the out-of-plane magnetic field, $B$.
The measurements reveal the dependence $T_1 \propto B^{-5}$, which is characteristic of a relaxation channel dominated by phonon-assisted spin flips driven by either a linear spin-orbit coupling or by a cubic Dresselhaus spin-orbit coupling (this dependence is, however, inconsistent with a cubic Rashba spin-orbit coupling).
The observation that the cubic Dresselhaus spin-orbit coupling can be neglected when calculating heavy-hole spin splittings (see Fig.~\ref{fig:ss}) provides some evidence that it can also be neglected in the context of spin relaxation.
Moreover, because the linear Dresselhaus spin-orbit coupling dominates the linear Rashba spin-orbit coupling (see Appendix \ref{app:Rashba}), it is plausible that the former is responsible for $T_1$ in Ref.~\cite{bogan2019single}.
Provided the linear Dresselhaus term ($\beta$) controls $T_1$ in this experiment, we can use the measured value as a quantitative check on our prediction for the value of $\beta$ shown in Fig.~\ref{fig:dress}.

In the limit of a weak out-of-plane magnetic field $B$ (hole Zeeman energy, $g\mu_BB$, small compared to the orbital level spacings $\epsilon_x$, $\epsilon_y$, for an elliptical dot) and for coupling to piezoelectric phonons (which dominates for weak magnetic fields), the relaxation rate due to phonon-assisted spin flips driven by linear Dresselhaus spin-orbit coupling is given by \cite{camenzind2018hyperfine}
\begin{equation}\label{eq:T1EllipticalDot}
    \frac{1}{T_1} \simeq \frac{4 (e h_{14} \beta)^2}{105 \pi \hbar^4 \rho c_t^5}\left(1 + \frac{3c_t^5}{4c_l^5}\right) (g\mu_B B)^5 \left(\frac{1}{\epsilon_x^4}+\frac{1}{\epsilon_y^4}\right).
\end{equation}
In this equation, $h_{14} = 1.4 \times 10^9\, \mathrm{V}/\mathrm{m}$ is the piezoelectric potential, $\rho = 5300\,\mathrm{kg}/\mathrm{m}^3$ is the material (GaAs) density, $c_t = 3350\,\mathrm{m}/\mathrm{s}$ is the transverse acoustic phonon velocity,  $c_l = 4730\,\mathrm{m}/\mathrm{s}$ is the longitudinal acoustic phonon velocity, $g=1.35$ is the heavy-hole out-of-plane $g$ factor \cite{bogan2019single}, and $\epsilon_x$, $\epsilon_y$ are the lowest two orbital level spacings for a quantum dot defined by parabolic confinement along the $x-$ and $y-$directions. 

The measurements of Ref.~\cite{bogan2019single} yield $T_1 = (2.5 \, \mu\mathrm{s}\cdot \mathrm{T}^5) B^{-5}$.
Using the range of coefficients presented in Fig.~\ref{fig:dress}, $\beta \simeq 0.35$-$0.40 \, \mathrm{meV}\, \mathrm{nm}$, and for an anisotropic quantum dot with orbital level spacings $\epsilon_x \simeq \epsilon_y/3\simeq0.3\,\mathrm{meV}$ (see the supplementary material of Ref.~\cite{bogan2017consequences}),\footnote{The device investigated in Ref.~\cite{bogan2017consequences} was fabricated with the same procedure as the device studied in Ref.~\cite{bogan2019single}.
In the supplementary material of Ref.~\cite{bogan2017consequences}, the quantum-dot spectrum is calculated based on a theoretical model and simulations of the quantum-dot device.
The orbital level spacings, $\epsilon_x$ and $\epsilon_y$, were extracted from this spectrum [Fig.~1(a) of the supplementary material of Ref.~\cite{bogan2017consequences} at $B=0$].} we obtain $T_1 \simeq (1.7$-$2.2 \, \mu\mathrm{s} \cdot \mathrm{T}^5) B^{-5}$, remarkably close to the measured value.\footnote{The range of values reported here would overlap with the measured value if, e.g., the true level spacing were larger than the estimated value ($\epsilon_x\simeq 0.3\,\mathrm{meV}$) by only 3\%.  Moreover, the hole-spin Zeeman splitting is comparable to the orbital level spacing for the range of magnetic fields ($B\simeq 0.5\,\mathrm{T}-1.5\,\mathrm{T}$) studied in Ref.~\cite{bogan2019single}.  For this range of magnetic fields, there could be a substantial (order unity) correction to Eq.~\eqref{eq:T1EllipticalDot}.}
This level of agreement suggests that the model presented in Sec.~\ref{sec:tripot}, which includes the dipolar spin-orbit coupling, may indeed give a highly accurate value for the linear Dresselhaus spin-orbit coupling.


\subsection{Comments on the model}
The model presented in this section attempts to capture the main features of the spin-orbit coupling for GaAs heavy holes confined to an asymmetric quantum well. 
Earlier works have calculated the heavy-hole spin splitting accounting for many-body effects by using a potential calculated self-consistently under the Hartree approximation (accounting for charges in both the inversion and depletion layers of an AlGaAs-GaAs heterojunction) \cite{ekenberg1984calculation, broido1985effective, bangert1985self}.
However, in these studies, bulk-inversion asymmetry was neglected.
More recently, heavy-hole spin splittings have been computed using wavefunctions obtained from a variational solution to the Poisson and Schr\"odinger equations \cite{marcellina2017spin}.
This procedure was shown to give results consistent with experiments, even in the parameter regime where the perturbation theory that projects the full Hamiltonian onto the lowest heavy-hole subband breaks down.
While the theory presented in Ref.~\cite{marcellina2017spin} includes bulk-inversion asymmetry by considering the terms ${\cal{H}}_1$ and ${\cal{H}}_3$, ${\cal{H}}_E$ is neglected. 

The potential considered in this paper is a triangular well [$U(z) = e E_zz$] which we take to represent the total effective potential experienced by holes at the heterointerface.
Although we have not fully accounted for many-body effects, the benefit of using an analytic form for the potential is that the spin-orbit couplings and spin splitting can be investigated analytically. 
Other works have adopted a similar approach.
For example, in Ref.~\cite{goldoni1995hole}, the spin-orbit coupling has also been calculated starting from the Luttinger Hamiltonian. 
However, in this reference terms involving bulk-inversion asymmetry were neglected (${\cal{H}}_1, {\cal{H}}_3,{\cal{H}}_E \rightarrow 0$).
If bulk inversion asymmetry is neglected in our model, the linear spin-orbit coupling vanishes and we find only a cubic Rashba heavy-hole spin-orbit coupling with coefficient $\gamma_1^R$, consistent with the results of Ref.~\cite{goldoni1995hole}.
In Ref.~\cite{wenk2016conserved}, the heavy-hole spin-orbit couplings were calculated using a similar procedure to that described in Sec.~\ref{sec:tripot}: The Schrieffer-Wolff transformation was applied to $\hat{H}$ to obtain the effective heavy-hole Hamiltonian $\hat{H}_{\mathrm{HH}}$.
Although bulk-inversion asymmetry was included in Ref.~\cite{wenk2016conserved} through the spin-orbit parameters $\beta_{C_k}$ and $\beta_{b1}$, the parameters $\beta_\chi,\beta_0$, and $\beta_{b2}$ were neglected.  In the present case, we find there is a range of electric field $E_z$ for which $\beta_\chi,\beta_0$, and $\beta_{b2}$ give the dominant contribution to the linear Dresselhaus spin-orbit coupling (and are all of the same order, see Figs.~\ref{fig:dress}, \ref{fig:ssvanishingchi}).

While the Hamiltonian matrix $\mathcal{H}$ [Eq.~\eqref{eq:fullH}] includes terms that describe bulk-inversion asymmetry, certain terms that describe structure-inversion asymmetry are neglected (these are the $k$-linear terms listed in Table 6.5 of Ref.~\cite{winkler2003spin}).
The dominant term among those that have been neglected has a coefficient $r_{41}$ (see Sec.~6.3.3 of Ref.~\cite{winkler2003spin} and Ref.~\cite{wenk2016conserved}) and off-diagonal matrix elements $\sim r_{41}E_z k_{\parallel}$.
This term couples the same valence-band states as the term with coefficient $b_{41}$ in Eq.~\eqref{eq:H3}, and which has off-diagonal matrix elements $\sim b_{41}\eta_2 k_{\parallel}$.
For GaAs, we find that $ |r_{41} E_z | < |b_{41}  \eta_2 |$ for the range $10^6\,\mathrm{V/m}\le |E_z|\lesssim 10^7\,\mathrm{V/m}$ considered here. 
Because terms that depend on $b_{41}$ are neglected (they lead to spin-orbit couplings that are smaller than those considered in Secs.~\ref{sec:cubic}~and~\ref{sec:Dresselhaus},
see Appendices~\ref{app:cubic}~and~\ref{app:Rashba} for justification), this justifies neglecting the contributions described above in the present work. 

Finally, we also note that in the model described here, $E_z$ provides the confinement that lifts the heavy-hole/light-hole degeneracy. 
Therefore, as $|E_z|$ decreases, $|\Delta_{\mathrm{HL}}|$ decreases.
A smaller splitting, $|\Delta_{\mathrm{HL}}|$, leads to larger corrections to the perturbation theory used to project ${\hat{H}}$ onto the two-dimensional heavy-hole subspace.
This perturbation theory has a small parameter $\varepsilon_{\mathrm{od}}/\Delta_{\mathrm{HL}}$, where $\varepsilon_{\mathrm{od}}$ represents an off-diagonal element of ${\cal{H}}$ (the magnitude of $\varepsilon_{\mathrm{od}}$ increases with $k_{\parallel}$).
For $E_z = - 10^6\,\mathrm{V}/\mathrm{m}$ and $k_{\parallel} \lesssim 0.1 \, \mathrm{nm}^{-1}$, we find $| \varepsilon_{\mathrm{od}}/\Delta_{\mathrm{HL}} | \lesssim 0.1$ for all off-diagonal elements $\varepsilon_{\mathrm{od}}$.  A breakdown of perturbation theory for $k_{\parallel} \gtrsim 0.1 \, \mathrm{nm}^{-1}$ is consistent with the results shown in Fig.~\ref{fig:MBSS}. See also Appendix \ref{app:higherbands} for further discussion of the validity of the perturbation theory.

\section{Conclusions}\label{sec:conclusion}
We have extended $\bm{k}\cdot\bm{p}$ theory to account for interband coupling generated by the triangular-well confining potential at a heterointerface. 
In subspaces where the basis states transform according to the $\Gamma_8$ representation of the tetrahedral double group (e.g., the heavy-hole and light-hole states at $k=0$ in a III-V semiconductor), this coupling can be parameterized by a single material-dependent parameter $\chi$.
Here, we have focused on the valence band of GaAs, but an equivalent analysis could be applied to other materials/bands.
Using the Kohn-Sham orbitals from an all-electron density functional theory calculation, we find $\chi = 0.2$ for GaAs.
This value for $\chi$ leads to a transition dipole of $ea_B\chi \simeq 0.5\, \mathrm{D}$, only a factor of $\sim 2$ smaller than that of a hydrogenic 1$s$ to 2$p$ transition.
Rabi-frequency measurements for the light-hole to heavy-hole transition in GaAs would allow for $\chi$ to be established experimentally.
The finite value of $\chi$ found here may be important for understanding electric-dipole spin resonance (EDSR) for GaAs hole-spin qubits with heavy-hole/light-hole mixing. 
This EDSR mechanism may even be present in group IV semiconductor nanostructures (silicon, germanium, \dots) that are sufficiently strained to significantly break inversion symmetry on the scale of the lattice.  

The finite value of $\chi\ne 0$ in III-V semiconductors (or in group IV materials with broken inversion symmetry due, e.g., to strain) leads to a new form of spin-orbit coupling: the dipolar spin-orbit coupling.
Because the dipolar spin-orbit coupling has Dresselhaus symmetry, it may be relevant to the measurements of Refs.~\cite{wang2016anisotropic,bogan2019single}, which show experimental evidence of Dresselhaus spin-orbit coupling.
More generally, a better understanding of the pseudospin-electric coupling may explain spin coherence/relaxation ($T_2^*/T_1$) times, spin-electric coupling for cavity-QED, electric-dipole spin resonance, and spin non-conserving tunneling in double quantum dot systems. 

A central observation of this paper is that the electric-dipole term $\mathcal{H}_{E}$ results in important physical consequences.  This term leads, e.g., to a modified spin splitting for heavy holes in a two-dimensional hole gas or to driven Rabi oscillations between heavy holes and light holes under the influence of a time-dependent electric field. The electric-dipole term does not vanish at $\bm{k}=0$, unlike all other inter-band terms in conventional $\bm{k}\cdot\bm{p}$ theory (within the envelope-function approximation).  Thus, long-wavelength (small $k$) properties of holes will generally be influenced by $\mathcal{H}_{E}$, even for $k\to0$, when the standard envelope-function approximation is expected to be accurate.

 \acknowledgments

The authors are grateful to R.~Winkler for helpful discussions.  WAC and PP acknowledge support from NSERC, FRQNT, and the Gordon Godfrey Bequest. DC was supported by the Australian Research Council Centre of Excellence in Future Low-Energy Electronics Technologies (project CE170100039) funded by the Australian Government. SC acknowledges support from the National Key Research and Development Program of China (Grant No. 2016YFA0301200), NSFC (Grant No. 11974040), and NSAF (Grant No. U1930402).

\appendix

\section{Calculating material parameters}\label{app:conv}

To calculate the $\bm{k} \cdot \bm{p}$, Zeeman-Hamiltonian, and position-operator matrix elements, we start by calculating the Kohn-Sham orbitals from the necessary bands at the $\Gamma$ point of GaAs.
Then, using the group-theoretic projection operators (see Ref.~\cite{dresselhaus2008} and Appendix C of Ref.~\cite{philippopoulos2020first}), we project these orbitals onto states with the appropriate symmetry.
For example, one $\bm{k}\cdot\bm{p}$ matrix element is \cite{winkler2003spin}
\begin{equation}\label{eq:P}
P = \frac{\hbar}{m}\bra{S}p_x\ket{X}, 
\end{equation}
where $\ket{S}$ is the $s$-like conduction-band $\bm{k} = \bm{0}$ Bloch function and $\ket{X}$ is the $p$-like valence-band $\bm{k} = \bm{0}$ Bloch function that transforms like the coordinate $x$ under the symmetry operations of the crystal. 
To evaluate the matrix element, Eq.~\eqref{eq:P}, we first calculate the $\Gamma$-point Kohn-Sham orbitals for the conduction and valence bands of GaAs.
Because the top of the valence band of GaAs is fourfold degenerate, a general valence-band Kohn-Sham orbital will be a linear combination of all four states (two heavy-hole and two light-hole states).
Using group-theoretic projection operators \cite{dresselhaus2008, philippopoulos2020first,philippopoulosgithub}, we project the general valence-band Kohn-Sham orbital onto the orbital that transforms like $x$ to obtain $\ket{X}$.  
Because the GaAs conduction band is $s$-like (isotropic), the only $\Gamma$-point conduction-band Kohn-Sham orbital is $\ket{S}$ and no projection is necessary in this case.  
Once we obtain $\ket{S}$ and $\ket{X}$, we evaluate $P$ (see Table~\ref{tab:kp}). 
    We apply a similar procedure to evaluate the other $\bm{k}\cdot\bm{p}$ matrix elements, $P'$ and $Q$ [$P$, $P'$, and $Q$ are defined in Eqs.~(3.3a), (3.3b) and (3.3c) of Ref.~\cite{winkler2003spin}], as well as $\kappa$, $q$, and $\chi$ (see Sec.~\ref{sec:fpkp}). The code used to implement these projections has been made freely available \cite{philippopoulosgithub}.

For all the parameters listed in Table \ref{tab:kp}, we have used the ``very high quality'' parameter set of the {\sc elk} code (\texttt{vhighq} set to \texttt{.true.} in the input file) \cite{elk} to calculate the Kohn-Sham orbitals. 
This set of parameters was shown to yield precise results (which have converged with respect to variation in multiple parameters to within 2\% of their asymptotic values, see Ref.~\cite{philippopoulos2020first}) when calculating the hyperfine couplings in GaAs and silicon \cite{philippopoulos2020first}. 

\section{Solution to triangular confining potential}\label{app:solutionstri}
Eq. \eqref{eq:SE} can be solved by the envelope functions 
\begin{equation}\label{eq:envfz}
 F_{\nu}^n(z)=C_{\nu} \mathrm{Ai}\left(\left[\frac{2m_{\nu}}{\hbar^2e^2E_z^2}\right]^{1/3}\left[e|E_z|z- \epsilon_{\nu}^n(z)\right]\right),
\end{equation}
where $C_{\nu}$ is a normalizing constant, and $\mathrm{Ai}(z)$ is an Airy function. The eigenenergies are given by 
\begin{equation}\label{eq:energy}
\epsilon_{\nu}^n=-\left(\frac{\hbar^2e^2E_z^2}{2m_{{\nu}}}\right)^{1/3}a_n,
\end{equation}
where $a_n$ is the $n^{\mathrm{th}}$ zero of $\mathrm{Ai}(z)$. If we define
\begin{equation}
\Lambda_{\nu}(z)=\left(\frac{2m_{\nu}e|E_z|}{\hbar^2}\right)^{1/3}\left(z-\frac{ \epsilon_{\nu}^n(z)}{e|E_z|}\right),
\end{equation}
we can write $C_{\nu}$ as
\begin{equation}
C_{\nu}=\left[\frac{\left(2m_{\nu}e|E_z|/\hbar^2\right)^{1/3}}{\mathrm{Ai}'(\Lambda_{\nu}(0))-\Lambda_{\nu}(0)\mathrm{Ai}^2(\Lambda_{\nu}(0))}\right]^{1/2}.
\end{equation}

\section{Contributions to the heavy-hole spin splitting from higher subbands}\label{app:higherbands}

\begin{figure}
\centering
\includegraphics[width=\columnwidth]{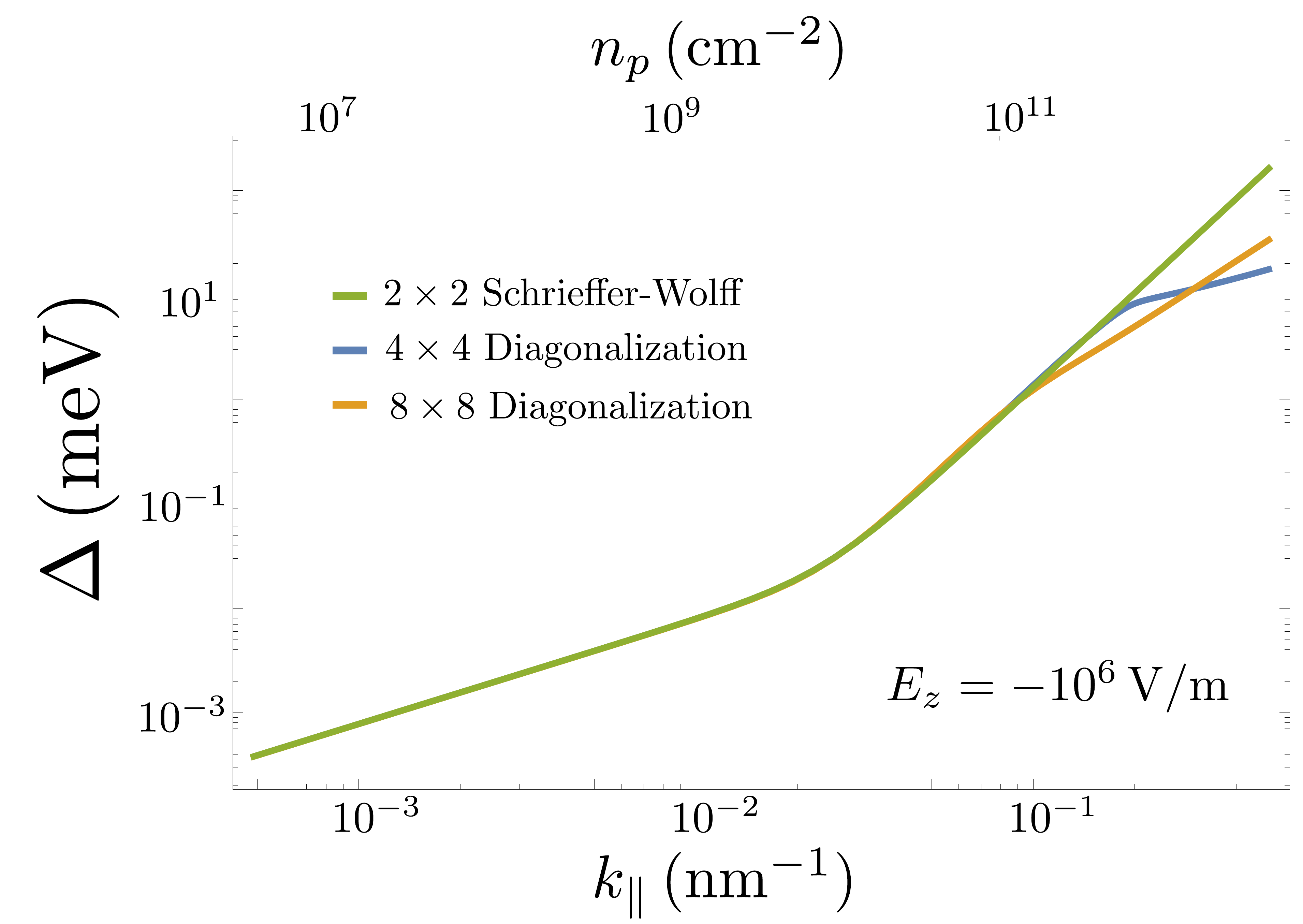}
\caption{Heavy-hole spin splitting in GaAs computed by diagonalizing the effective $2 \times 2$ Hamiltonian obtained from the Schrieffer-Wolff transformation, as described in the main text (in green), by diagonalizing the $4 \times 4$ Hamiltonian in the lowest valence-band subband (in blue), and by diagonalizing the $8 \times 8$ Hamiltonian in the two lowest valence-band subbands (in yellow). 
The top horizontal axis indicates the hole sheet density,  $n_p = k_{\parallel}^2 /(2\pi)$, associated with a Fermi wavevector, $k_{F}=k_{\parallel}$.
All calculations were performed for a triangular well with electric field $E_z = - 10^6 \, \mathrm{V}/ \mathrm{m}$ with material parameters used (for GaAs) given in Table \ref{tab:params}.
}\label{fig:MBSS}
\end{figure}

In the approach described in the main text, the full Hamiltonian ${\cal{H}}$ [Eq.~\eqref{eq:fullH}] is projected onto the lowest subband to obtain the four-dimensional Hamiltonian, $\hat{H}$. 
$\hat{H}$ is then projected onto the heavy-hole subspace using perturbation theory (Schrieffer-Wolff transformation) to obtain an effective two-dimensional Hamiltonian, $\hat{H}_{\mathrm{HH}}$ [Eq.~\eqref{eq:2x2Hamiltonian}]. 
To verify the spin-splitting obtained from the Schrieffer-Wolff procedure (green line in Fig.~\ref{fig:MBSS}), we have numerically diagonalized the four-dimensional Hamiltonian arising from the lowest subband, $\hat{H}$, resulting in the spin splitting shown in the blue line of Fig.~\ref{fig:MBSS}.
To address the effect of the first excited subband, we have also numerically diagonalized the eight-dimensional Hamiltonian, giving the yellow line shown in Fig.~\ref{fig:MBSS}.
From Fig.~\ref{fig:MBSS}, perturbation theory breaks down at large $k_\parallel$ ($k_\parallel\gtrsim 10^{-1} \, \mathrm{nm}^{-1}$).  For the analysis presented here to give quantitatively accurate results, we therefore require $k_\parallel\lesssim 10^{-1} \, \mathrm{nm}^{-1}$.  For a two-dimensional hole gas with Fermi wavevector $k_\mathrm{F}=k_\parallel$, this implies a low sheet density, $n_p\lesssim 10^{-11}\,\mathrm{cm}$.

\section{Linear Rashba spin-orbit coupling}\label{app:Rashba}

\begin{figure}
\includegraphics[width=\columnwidth]{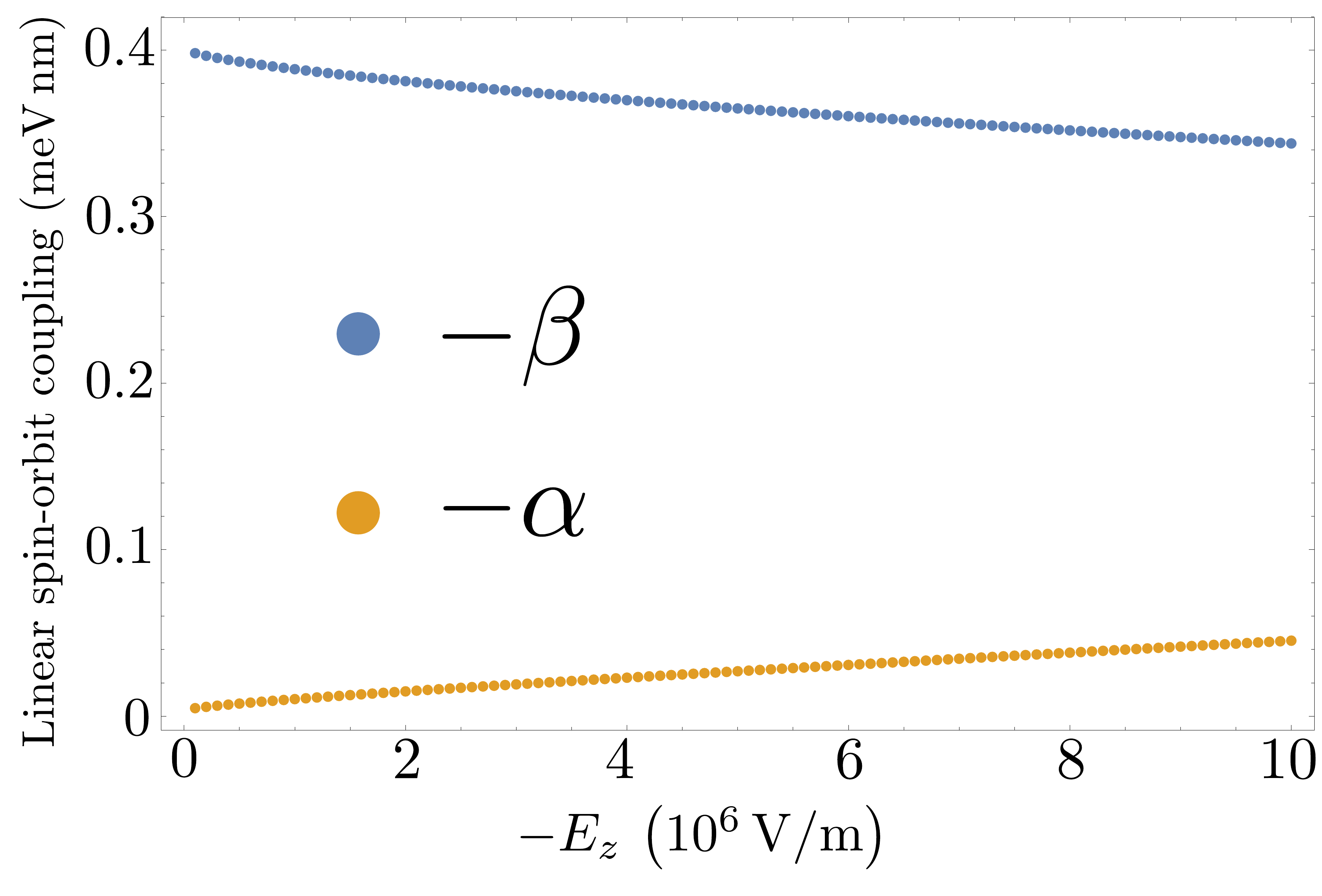}
\caption[rashba]{Linear spin-orbit coupling coefficients. The magnitude of the Dresselhaus coefficient (in blue), $\beta$, is larger than the magnitude of the Rashba coefficient (in yellow), $\alpha$, for all values of $E_z$ considered. 
The values for the material parameters (for GaAs) are listed in Table \ref{tab:params}. }\label{fig:rashba}
\end{figure}

In addition to the linear Dresselhaus spin-orbit coupling discussed in Sec.~\ref{sec:Dresselhaus}, a linear Rashba  spin-orbit coupling can be derived from the same model: 
\begin{equation}\label{eq:Rashba}
\hat{H}_R = i \alpha 
	\left(k_+\sigma_+-k_-\sigma_-\right),
\end{equation}
where $\alpha$ is the linear Rashba spin-orbit coupling coefficient. 
This coefficient depends on the strength of the electric field, $E_z$, as well as the material-specific parameters $\chi$, $C_k$, $b_{41}$, $b_{42}$, $b_{51}$, $b_{52}$ and $\gamma_2$.
In Fig.~\ref{fig:rashba} we compare the value of $\alpha$ to the linear Dresselhaus coefficient, $\beta$.
We find that (for the values of $E_z$ considered) $|\alpha| \lesssim  10\% |\beta|$.
For simplicity, We neglect the linear Rashba spin-orbit coupling in the analysis presented in the main text.  
However, we note that the the linear Rashba coefficient could be calculated and included in the analysis.
The procedure to do so would be similar to that outlined in Sec.~\ref{sec:tripot}: project the Hamiltonian ${\cal{H}}$ onto the lowest subband subspace to obtain ${\hat{H}}$, perform the Schrieffer-Wolff transformation on the Hamiltonian ${\hat{H}}$ to project it onto the heavy-hole subspace, and then collect the terms linear in $k_{\parallel}$ that possess Rashba symmetry [Eq.~\eqref{eq:Rashba}].

\begin{figure}
\includegraphics[width=\columnwidth]{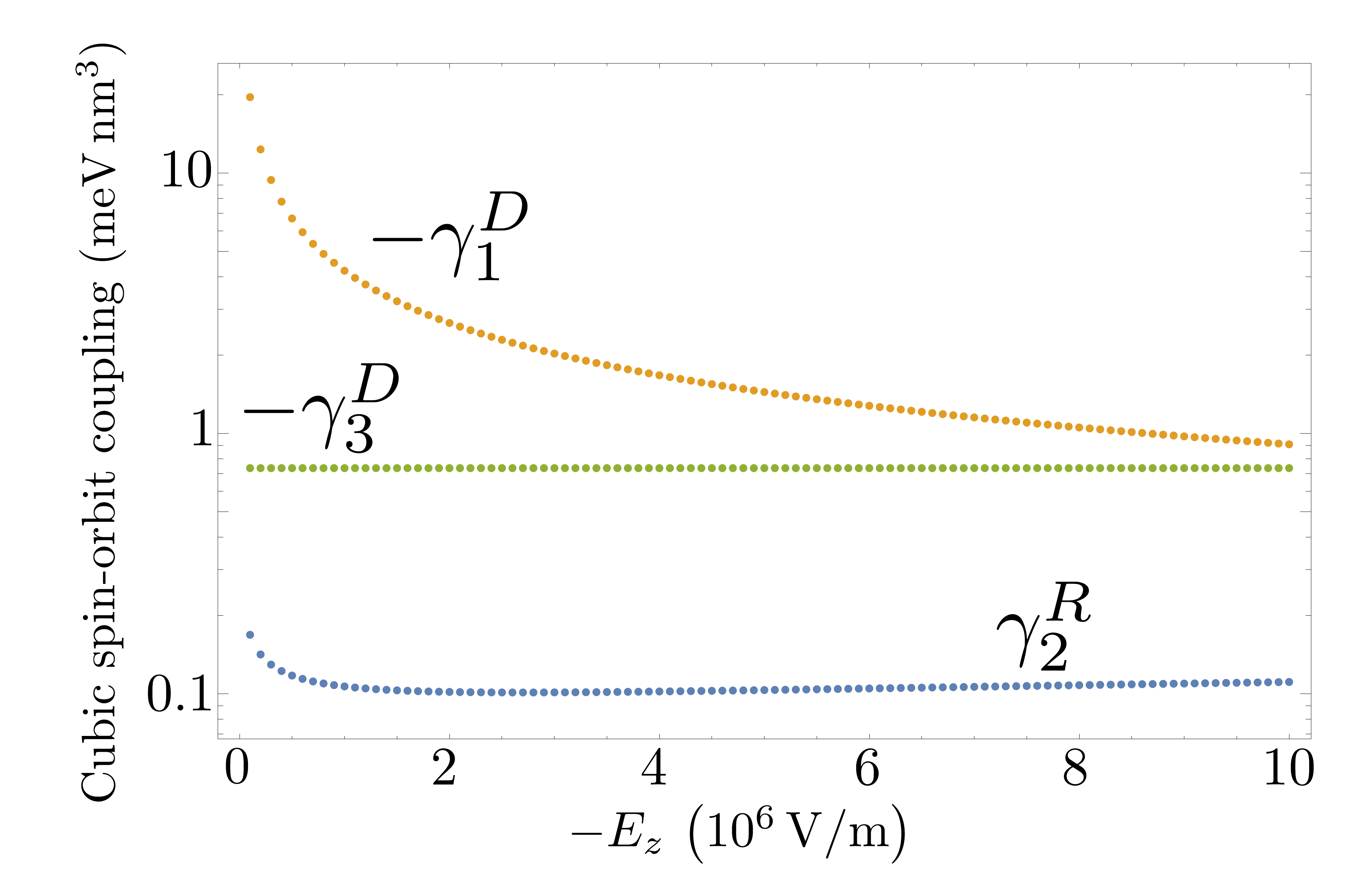}
\caption[rashba]{Cubic spin-orbit coupling coefficients $-\gamma_1^D$ (in yellow), $-\gamma_3^D$ (in green), and $\gamma_2^R$ (in blue) as a function of electric field, $E_z$.
$|\gamma_1^D|$, is orders of magnitude smaller than the dominant cubic spin-orbit coupling term characterized by $\gamma_1^R$ (see Sec.~\ref{sec:cubic}), however it is larger (in magnitude) than both $\gamma_3^D$ and $\gamma_2^R$, for the values of $E_z$ considered here. The values for the material parameters (for GaAs) are listed in Table \ref{tab:params}. }\label{fig:cubiccorr}
\end{figure}

\section{Additional cubic spin-orbit coupling}\label{app:cubic}

In addition to the cubic spin-orbit couplings discussed in Sec.~\ref{sec:cubic} ($\gamma_1^R$, $\gamma_1^D$, and $\gamma_2^D$) there are extra small terms that we discuss here for completeness.
According to the model presented in Sec.~\ref{sec:tripot}, the cubic Rashba spin-orbit coupling coefficient is given by,
\begin{equation}
\gamma^R = \gamma_1^R+\gamma_2^R,
\end{equation}
where $\gamma_1^R$ is given by Eq.~\eqref{eq:gammaR1} of the main text and
\begin{align}
&\gamma^R_2 = \frac{4b_{41} + 7b_{42} - 2b_{51} + 2b_{52}}{32\Delta_{\mathrm{HL}}} \xi \\
& \times \left(3b_{52}\eta_3 - 2\sqrt{3}C_k\eta_1 - \sqrt{3} E_z ea_B \chi \xi\right) \nonumber.
\end{align}
Similarly, the Dresselhaus spin-orbit coupling has an additional contribution:
\begin{equation}
\gamma^D = \gamma_1^D+\gamma_2^D+\gamma_3^D,
\end{equation}
where $\gamma_1^D$ is given by Eq.~\eqref{eq:gammaD1} of the main text, $\gamma_2^D$ is given by Eq.~\eqref{eq:gammaD2} of the main text, and
\begin{equation}
\gamma^D_3 = \frac{3}{16}(b_{51} + 3b_{52} - b_{42}).
\end{equation}  
As shown in Fig.~\ref{fig:cubic}, for the range of electric fields considered, the cubic spin-orbit coupling parameters satisfy $|\gamma_1^R| \gg |\gamma^D_2|>|\gamma_1^D|$ .
The additional terms contributing to $\gamma^R$ and $\gamma^D$ ($\gamma_2^R$ and $\gamma^D_3$) are even smaller (in magnitude) than $\gamma_1^D$ (see Fig.~\ref{fig:cubiccorr}) over the range of electric fields considered and are therefore neglected throughout the main text.


\bibliographystyle{apsrev4-2}
\bibliography{SOcoupling}

\begin{thebibliography}{38}%
\makeatletter
\providecommand \@ifxundefined [1]{%
 \@ifx{#1\undefined}
}%
\providecommand \@ifnum [1]{%
 \ifnum #1\expandafter \@firstoftwo
 \else \expandafter \@secondoftwo
 \fi
}%
\providecommand \@ifx [1]{%
 \ifx #1\expandafter \@firstoftwo
 \else \expandafter \@secondoftwo
 \fi
}%
\providecommand \natexlab [1]{#1}%
\providecommand \enquote  [1]{``#1''}%
\providecommand \bibnamefont  [1]{#1}%
\providecommand \bibfnamefont [1]{#1}%
\providecommand \citenamefont [1]{#1}%
\providecommand \href@noop [0]{\@secondoftwo}%
\providecommand \href [0]{\begingroup \@sanitize@url \@href}%
\providecommand \@href[1]{\@@startlink{#1}\@@href}%
\providecommand \@@href[1]{\endgroup#1\@@endlink}%
\providecommand \@sanitize@url [0]{\catcode `\\12\catcode `\$12\catcode
  `\&12\catcode `\#12\catcode `\^12\catcode `\_12\catcode `\%12\relax}%
\providecommand \@@startlink[1]{}%
\providecommand \@@endlink[0]{}%
\providecommand \url  [0]{\begingroup\@sanitize@url \@url }%
\providecommand \@url [1]{\endgroup\@href {#1}{\urlprefix }}%
\providecommand \urlprefix  [0]{URL }%
\providecommand \Eprint [0]{\href }%
\providecommand \doibase [0]{https://doi.org/}%
\providecommand \selectlanguage [0]{\@gobble}%
\providecommand \bibinfo  [0]{\@secondoftwo}%
\providecommand \bibfield  [0]{\@secondoftwo}%
\providecommand \translation [1]{[#1]}%
\providecommand \BibitemOpen [0]{}%
\providecommand \bibitemStop [0]{}%
\providecommand \bibitemNoStop [0]{.\EOS\space}%
\providecommand \EOS [0]{\spacefactor3000\relax}%
\providecommand \BibitemShut  [1]{\csname bibitem#1\endcsname}%
\let\auto@bib@innerbib\@empty
\bibitem [{\citenamefont {Winkler}(2003)}]{winkler2003spin}%
  \BibitemOpen
  \bibfield  {author} {\bibinfo {author} {\bibfnamefont {R.}~\bibnamefont
  {Winkler}},\ }\href@noop {} {\emph {\bibinfo {title} {Spin-Orbit Coupling
  Effects in Two-Dimensional Electron and Hole Systems}}}\ (\bibinfo
  {publisher} {Springer},\ \bibinfo {year} {2003})\BibitemShut {NoStop}%
\bibitem [{\citenamefont {Lew Yan~Voon}\ and\ \citenamefont
  {Willatzen}(2009)}]{voon2009kp}%
  \BibitemOpen
  \bibfield  {author} {\bibinfo {author} {\bibfnamefont {L.~C.}\ \bibnamefont
  {Lew Yan~Voon}}\ and\ \bibinfo {author} {\bibfnamefont {M.}~\bibnamefont
  {Willatzen}},\ }\href@noop {} {\emph {\bibinfo {title} {The $\bm{k}\cdot
  \bm{p}$ method: electronic properties of semiconductors}}}\ (\bibinfo
  {publisher} {Springer},\ \bibinfo {year} {2009})\BibitemShut {NoStop}%
\bibitem [{\citenamefont {Bychkov}\ and\ \citenamefont
  {Rashba}(1984)}]{bychkov1984oscillatory}%
  \BibitemOpen
  \bibfield  {author} {\bibinfo {author} {\bibfnamefont {Y.~A.}\ \bibnamefont
  {Bychkov}}\ and\ \bibinfo {author} {\bibfnamefont {E.~I.}\ \bibnamefont
  {Rashba}},\ }\href {https://doi.org/10.1088/0022-3719/17/33/015} {\bibfield
  {journal} {\bibinfo  {journal} {J.~Phys.~C Solid State Phys.}\ }\textbf
  {\bibinfo {volume} {17}},\ \bibinfo {pages} {6039} (\bibinfo {year}
  {1984})}\BibitemShut {NoStop}%
\bibitem [{\citenamefont {Dresselhaus}(1955)}]{dresselhaus1955spin}%
  \BibitemOpen
  \bibfield  {author} {\bibinfo {author} {\bibfnamefont {G.}~\bibnamefont
  {Dresselhaus}},\ }\href {https://doi.org/10.1103/PhysRev.100.580} {\bibfield
  {journal} {\bibinfo  {journal} {Phys. Rev.}\ }\textbf {\bibinfo {volume}
  {100}},\ \bibinfo {pages} {580} (\bibinfo {year} {1955})}\BibitemShut
  {NoStop}%
\bibitem [{\citenamefont {Bir}\ \emph {et~al.}(1963)\citenamefont {Bir},
  \citenamefont {Butikov},\ and\ \citenamefont {Pikus}}]{bir1963spin}%
  \BibitemOpen
  \bibfield  {author} {\bibinfo {author} {\bibfnamefont {G.}~\bibnamefont
  {Bir}}, \bibinfo {author} {\bibfnamefont {E.}~\bibnamefont {Butikov}},\ and\
  \bibinfo {author} {\bibfnamefont {G.}~\bibnamefont {Pikus}},\ }\href@noop {}
  {\bibfield  {journal} {\bibinfo  {journal} {J.~Phys.~Chem.~Solids}\ }\textbf
  {\bibinfo {volume} {24}},\ \bibinfo {pages} {1475} (\bibinfo {year}
  {1963})}\BibitemShut {NoStop}%
\bibitem [{\citenamefont {K{\"o}pf}\ and\ \citenamefont
  {Lassmann}(1992)}]{kopf1992linear}%
  \BibitemOpen
  \bibfield  {author} {\bibinfo {author} {\bibfnamefont {A.}~\bibnamefont
  {K{\"o}pf}}\ and\ \bibinfo {author} {\bibfnamefont {K.}~\bibnamefont
  {Lassmann}},\ }\href@noop {} {\bibfield  {journal} {\bibinfo  {journal}
  {\prl}\ }\textbf {\bibinfo {volume} {69}},\ \bibinfo {pages} {1580} (\bibinfo
  {year} {1992})}\BibitemShut {NoStop}%
\bibitem [{\citenamefont {Abadillo-Uriel}\ \emph {et~al.}(2018)\citenamefont
  {Abadillo-Uriel}, \citenamefont {Salfi}, \citenamefont {Hu}, \citenamefont
  {Rogge}, \citenamefont {Calder{\'o}n},\ and\ \citenamefont
  {Culcer}}]{abadillo2018entanglement}%
  \BibitemOpen
  \bibfield  {author} {\bibinfo {author} {\bibfnamefont {J.~C.}\ \bibnamefont
  {Abadillo-Uriel}}, \bibinfo {author} {\bibfnamefont {J.}~\bibnamefont
  {Salfi}}, \bibinfo {author} {\bibfnamefont {X.}~\bibnamefont {Hu}}, \bibinfo
  {author} {\bibfnamefont {S.}~\bibnamefont {Rogge}}, \bibinfo {author}
  {\bibfnamefont {M.~J.}\ \bibnamefont {Calder{\'o}n}},\ and\ \bibinfo {author}
  {\bibfnamefont {D.}~\bibnamefont {Culcer}},\ }\href@noop {} {\bibfield
  {journal} {\bibinfo  {journal} {Appl.~Phys.~Lett.}\ }\textbf {\bibinfo
  {volume} {113}},\ \bibinfo {pages} {012102} (\bibinfo {year}
  {2018})}\BibitemShut {NoStop}%
\bibitem [{\citenamefont {Philippopoulos}\ \emph {et~al.}(2020)\citenamefont
  {Philippopoulos}, \citenamefont {Chesi},\ and\ \citenamefont
  {Coish}}]{philippopoulos2020first}%
  \BibitemOpen
  \bibfield  {author} {\bibinfo {author} {\bibfnamefont {P.}~\bibnamefont
  {Philippopoulos}}, \bibinfo {author} {\bibfnamefont {S.}~\bibnamefont
  {Chesi}},\ and\ \bibinfo {author} {\bibfnamefont {W.~A.}\ \bibnamefont
  {Coish}},\ }\href {https://doi.org/10.1103/PhysRevB.101.115302} {\bibfield
  {journal} {\bibinfo  {journal} {Phys. Rev. B}\ }\textbf {\bibinfo {volume}
  {101}},\ \bibinfo {pages} {115302} (\bibinfo {year} {2020})}\BibitemShut
  {NoStop}%
\bibitem [{\citenamefont {Wang}\ \emph {et~al.}(2016)\citenamefont {Wang},
  \citenamefont {Klochan}, \citenamefont {Hung}, \citenamefont {Culcer},
  \citenamefont {Farrer}, \citenamefont {Ritchie},\ and\ \citenamefont
  {Hamilton}}]{wang2016anisotropic}%
  \BibitemOpen
  \bibfield  {author} {\bibinfo {author} {\bibfnamefont {D.~Q.}\ \bibnamefont
  {Wang}}, \bibinfo {author} {\bibfnamefont {O.}~\bibnamefont {Klochan}},
  \bibinfo {author} {\bibfnamefont {J.-T.}\ \bibnamefont {Hung}}, \bibinfo
  {author} {\bibfnamefont {D.}~\bibnamefont {Culcer}}, \bibinfo {author}
  {\bibfnamefont {I.}~\bibnamefont {Farrer}}, \bibinfo {author} {\bibfnamefont
  {D.~A.}\ \bibnamefont {Ritchie}},\ and\ \bibinfo {author} {\bibfnamefont
  {A.~R.}\ \bibnamefont {Hamilton}},\ }\href@noop {} {\bibfield  {journal}
  {\bibinfo  {journal} {Nano~Lett.}\ }\textbf {\bibinfo {volume} {16}},\
  \bibinfo {pages} {7685} (\bibinfo {year} {2016})}\BibitemShut {NoStop}%
\bibitem [{\citenamefont {Bogan}\ \emph {et~al.}(2019)\citenamefont {Bogan},
  \citenamefont {Studenikin}, \citenamefont {Korkusinski}, \citenamefont
  {Gaudreau}, \citenamefont {Zawadzki}, \citenamefont {Sachrajda},
  \citenamefont {Tracy}, \citenamefont {Reno},\ and\ \citenamefont
  {Hargett}}]{bogan2019single}%
  \BibitemOpen
  \bibfield  {author} {\bibinfo {author} {\bibfnamefont {A.}~\bibnamefont
  {Bogan}}, \bibinfo {author} {\bibfnamefont {S.}~\bibnamefont {Studenikin}},
  \bibinfo {author} {\bibfnamefont {M.}~\bibnamefont {Korkusinski}}, \bibinfo
  {author} {\bibfnamefont {L.}~\bibnamefont {Gaudreau}}, \bibinfo {author}
  {\bibfnamefont {P.}~\bibnamefont {Zawadzki}}, \bibinfo {author}
  {\bibfnamefont {A.}~\bibnamefont {Sachrajda}}, \bibinfo {author}
  {\bibfnamefont {L.}~\bibnamefont {Tracy}}, \bibinfo {author} {\bibfnamefont
  {J.}~\bibnamefont {Reno}},\ and\ \bibinfo {author} {\bibfnamefont
  {T.}~\bibnamefont {Hargett}},\ }\href@noop {} {\bibfield  {journal} {\bibinfo
   {journal} {Commun.~Phys.}\ }\textbf {\bibinfo {volume} {2}},\ \bibinfo
  {pages} {17} (\bibinfo {year} {2019})}\BibitemShut {NoStop}%
\bibitem [{\citenamefont {Bulaev}\ and\ \citenamefont
  {Loss}(2005)}]{bulaev2005spin}%
  \BibitemOpen
  \bibfield  {author} {\bibinfo {author} {\bibfnamefont {D.~V.}\ \bibnamefont
  {Bulaev}}\ and\ \bibinfo {author} {\bibfnamefont {D.}~\bibnamefont {Loss}},\
  }\href {https://doi.org/10.1103/PhysRevLett.95.076805} {\bibfield  {journal}
  {\bibinfo  {journal} {Phys. Rev. Lett.}\ }\textbf {\bibinfo {volume} {95}},\
  \bibinfo {pages} {076805} (\bibinfo {year} {2005})}\BibitemShut {NoStop}%
\bibitem [{\citenamefont {Luttinger}\ and\ \citenamefont
  {Kohn}(1955)}]{luttinger1955motion}%
  \BibitemOpen
  \bibfield  {author} {\bibinfo {author} {\bibfnamefont {J.~M.}\ \bibnamefont
  {Luttinger}}\ and\ \bibinfo {author} {\bibfnamefont {W.}~\bibnamefont
  {Kohn}},\ }\href@noop {} {\bibfield  {journal} {\bibinfo  {journal} {Phys.
  Rev.}\ }\textbf {\bibinfo {volume} {97}},\ \bibinfo {pages} {869} (\bibinfo
  {year} {1955})}\BibitemShut {NoStop}%
\bibitem [{\citenamefont {Elder}\ \emph {et~al.}(2011)\citenamefont {Elder},
  \citenamefont {Ward}, \citenamefont {Zhang} \emph
  {et~al.}}]{elder2011double}%
  \BibitemOpen
  \bibfield  {author} {\bibinfo {author} {\bibfnamefont {W.~J.}\ \bibnamefont
  {Elder}}, \bibinfo {author} {\bibfnamefont {R.~M.}\ \bibnamefont {Ward}},
  \bibinfo {author} {\bibfnamefont {J.}~\bibnamefont {Zhang}}, \emph {et~al.},\
  }\href@noop {} {\bibfield  {journal} {\bibinfo  {journal} {\prb}\ }\textbf
  {\bibinfo {volume} {83}},\ \bibinfo {pages} {165210} (\bibinfo {year}
  {2011})}\BibitemShut {NoStop}%
\bibitem [{\citenamefont {Lew Yan~Voon}\ and\ \citenamefont
  {Ram-Mohan}(1993)}]{lewyanvoon1993tight}%
  \BibitemOpen
  \bibfield  {author} {\bibinfo {author} {\bibfnamefont {L.~C.}\ \bibnamefont
  {Lew Yan~Voon}}\ and\ \bibinfo {author} {\bibfnamefont {L.~R.}\ \bibnamefont
  {Ram-Mohan}},\ }\href {https://doi.org/10.1103/PhysRevB.47.15500} {\bibfield
  {journal} {\bibinfo  {journal} {Phys. Rev. B}\ }\textbf {\bibinfo {volume}
  {47}},\ \bibinfo {pages} {15500} (\bibinfo {year} {1993})}\BibitemShut
  {NoStop}%
\bibitem [{\citenamefont {Gorczyca}\ \emph {et~al.}(1991)\citenamefont
  {Gorczyca}, \citenamefont {Pfeffer},\ and\ \citenamefont
  {Zawadzki}}]{gorczyca1991pseudopotential}%
  \BibitemOpen
  \bibfield  {author} {\bibinfo {author} {\bibfnamefont {I.}~\bibnamefont
  {Gorczyca}}, \bibinfo {author} {\bibfnamefont {P.}~\bibnamefont {Pfeffer}},\
  and\ \bibinfo {author} {\bibfnamefont {W.}~\bibnamefont {Zawadzki}},\
  }\href@noop {} {\bibfield  {journal} {\bibinfo  {journal}
  {Semicond.~Sci.~Technol.}\ }\textbf {\bibinfo {volume} {6}},\ \bibinfo
  {pages} {963} (\bibinfo {year} {1991})}\BibitemShut {NoStop}%
\bibitem [{\citenamefont {Pugh}\ \emph {et~al.}(1999)\citenamefont {Pugh},
  \citenamefont {Dugdale}, \citenamefont {Brand},\ and\ \citenamefont
  {Abram}}]{pugh1999band}%
  \BibitemOpen
  \bibfield  {author} {\bibinfo {author} {\bibfnamefont {S.~K.}\ \bibnamefont
  {Pugh}}, \bibinfo {author} {\bibfnamefont {D.~J.}\ \bibnamefont {Dugdale}},
  \bibinfo {author} {\bibfnamefont {S.}~\bibnamefont {Brand}},\ and\ \bibinfo
  {author} {\bibfnamefont {R.~A.}\ \bibnamefont {Abram}},\ }\href@noop {}
  {\bibfield  {journal} {\bibinfo  {journal} {J. Appl. Phys.}\ }\textbf
  {\bibinfo {volume} {86}},\ \bibinfo {pages} {3768} (\bibinfo {year}
  {1999})}\BibitemShut {NoStop}%
\bibitem [{\citenamefont {Tan}\ \emph {et~al.}(2013)\citenamefont {Tan},
  \citenamefont {Povolotskyi}, \citenamefont {Kubis}, \citenamefont {He},
  \citenamefont {Jiang}, \citenamefont {Klimeck},\ and\ \citenamefont
  {Boykin}}]{tan2013empirical}%
  \BibitemOpen
  \bibfield  {author} {\bibinfo {author} {\bibfnamefont {Y.}~\bibnamefont
  {Tan}}, \bibinfo {author} {\bibfnamefont {M.}~\bibnamefont {Povolotskyi}},
  \bibinfo {author} {\bibfnamefont {T.}~\bibnamefont {Kubis}}, \bibinfo
  {author} {\bibfnamefont {Y.}~\bibnamefont {He}}, \bibinfo {author}
  {\bibfnamefont {Z.}~\bibnamefont {Jiang}}, \bibinfo {author} {\bibfnamefont
  {G.}~\bibnamefont {Klimeck}},\ and\ \bibinfo {author} {\bibfnamefont {T.~B.}\
  \bibnamefont {Boykin}},\ }\href@noop {} {\bibfield  {journal} {\bibinfo
  {journal} {J. Comput Electron.}\ }\textbf {\bibinfo {volume} {12}},\ \bibinfo
  {pages} {56} (\bibinfo {year} {2013})}\BibitemShut {NoStop}%
\bibitem [{\citenamefont {Dewhurst}\ \emph {et~al.}(2008)\citenamefont
  {Dewhurst}, \citenamefont {Sharma}, \citenamefont {Nordstrom}, \citenamefont
  {Cricchio}, \citenamefont {Bultmark}, \citenamefont {Granas},\ and\
  \citenamefont {Gross}}]{elk}%
  \BibitemOpen
  \bibfield  {author} {\bibinfo {author} {\bibfnamefont {J.~K.}\ \bibnamefont
  {Dewhurst}}, \bibinfo {author} {\bibfnamefont {S.}~\bibnamefont {Sharma}},
  \bibinfo {author} {\bibfnamefont {L.}~\bibnamefont {Nordstrom}}, \bibinfo
  {author} {\bibfnamefont {F.}~\bibnamefont {Cricchio}}, \bibinfo {author}
  {\bibfnamefont {F.}~\bibnamefont {Bultmark}}, \bibinfo {author}
  {\bibfnamefont {O.}~\bibnamefont {Granas}},\ and\ \bibinfo {author}
  {\bibfnamefont {E.~K.~U.}\ \bibnamefont {Gross}},\ }\href
  {http://elk.sourceforge.net/elk.pdf} {\bibinfo {title} {The {\sc elk} code
  manual version 2.3.22}} (\bibinfo {year} {2008})\BibitemShut {NoStop}%
\bibitem [{\citenamefont {http://elk.sourceforge.net/}()}]{elkwebsite}%
  \BibitemOpen
  \bibfield  {author} {\bibinfo {author} {\bibnamefont
  {http://elk.sourceforge.net/}},\ }\href@noop {} {\bibinfo {title} {The {\sc
  elk} code}}\BibitemShut {NoStop}%
\bibitem [{\citenamefont {Luttinger}(1956)}]{luttinger1956quantum}%
  \BibitemOpen
  \bibfield  {author} {\bibinfo {author} {\bibfnamefont {J.~M.}\ \bibnamefont
  {Luttinger}},\ }\href@noop {} {\bibfield  {journal} {\bibinfo  {journal}
  {Phys. Rev.}\ }\textbf {\bibinfo {volume} {102}},\ \bibinfo {pages} {1030}
  (\bibinfo {year} {1956})}\BibitemShut {NoStop}%
\bibitem [{\citenamefont {Winkler}(2004)}]{winkler2004spin}%
  \BibitemOpen
  \bibfield  {author} {\bibinfo {author} {\bibfnamefont {R.}~\bibnamefont
  {Winkler}},\ }\href {https://doi.org/10.1103/PhysRevB.70.125301} {\bibfield
  {journal} {\bibinfo  {journal} {Phys. Rev. B}\ }\textbf {\bibinfo {volume}
  {70}},\ \bibinfo {pages} {125301} (\bibinfo {year} {2004})}\BibitemShut
  {NoStop}%
\bibitem [{\citenamefont {Culcer}\ \emph {et~al.}(2006)\citenamefont {Culcer},
  \citenamefont {Lechner},\ and\ \citenamefont {Winkler}}]{culcer2006spin}%
  \BibitemOpen
  \bibfield  {author} {\bibinfo {author} {\bibfnamefont {D.}~\bibnamefont
  {Culcer}}, \bibinfo {author} {\bibfnamefont {C.}~\bibnamefont {Lechner}},\
  and\ \bibinfo {author} {\bibfnamefont {R.}~\bibnamefont {Winkler}},\ }\href
  {https://doi.org/10.1103/PhysRevLett.97.106601} {\bibfield  {journal}
  {\bibinfo  {journal} {Phys. Rev. Lett.}\ }\textbf {\bibinfo {volume} {97}},\
  \bibinfo {pages} {106601} (\bibinfo {year} {2006})}\BibitemShut {NoStop}%
\bibitem [{\citenamefont {Abragam}(1961)}]{abragam1961principlesp163}%
  \BibitemOpen
  \bibfield  {author} {\bibinfo {author} {\bibfnamefont {A.}~\bibnamefont
  {Abragam}},\ }\href@noop {} {\emph {\bibinfo {title} {The Principles of
  Nuclear Magnetism}}}\ (\bibinfo  {publisher} {Clarendon Press},\ \bibinfo
  {address} {Oxford, U.K.},\ \bibinfo {year} {1961})\ pp.\ \bibinfo {pages}
  {162 -- 164}\BibitemShut {NoStop}%
\bibitem [{\citenamefont {Perdew}(1985)}]{perdew1985density}%
  \BibitemOpen
  \bibfield  {author} {\bibinfo {author} {\bibfnamefont {J.~P.}\ \bibnamefont
  {Perdew}},\ }\href {https://doi.org/10.1002/qua.560280846} {\bibfield
  {journal} {\bibinfo  {journal} {Int. J. Quantum Chem.}\ }\textbf {\bibinfo
  {volume} {28}},\ \bibinfo {pages} {497} (\bibinfo {year} {1985})}\BibitemShut
  {NoStop}%
\bibitem [{\citenamefont {Cardona}\ \emph {et~al.}(1988)\citenamefont
  {Cardona}, \citenamefont {Christensen},\ and\ \citenamefont
  {Fasol}}]{cardona1988relativistic}%
  \BibitemOpen
  \bibfield  {author} {\bibinfo {author} {\bibfnamefont {M.}~\bibnamefont
  {Cardona}}, \bibinfo {author} {\bibfnamefont {N.~E.}\ \bibnamefont
  {Christensen}},\ and\ \bibinfo {author} {\bibfnamefont {G.}~\bibnamefont
  {Fasol}},\ }\href {https://doi.org/10.1103/PhysRevB.38.1806} {\bibfield
  {journal} {\bibinfo  {journal} {Phys. Rev. B}\ }\textbf {\bibinfo {volume}
  {38}},\ \bibinfo {pages} {1806} (\bibinfo {year} {1988})}\BibitemShut
  {NoStop}%
\bibitem [{\citenamefont {Schrieffer}\ and\ \citenamefont
  {Wolff}(1966)}]{schrieffer1966relation}%
  \BibitemOpen
  \bibfield  {author} {\bibinfo {author} {\bibfnamefont {J.~R.}\ \bibnamefont
  {Schrieffer}}\ and\ \bibinfo {author} {\bibfnamefont {P.~A.}\ \bibnamefont
  {Wolff}},\ }\href {https://doi.org/10.1103/PhysRev.149.491} {\bibfield
  {journal} {\bibinfo  {journal} {Phys. Rev.}\ }\textbf {\bibinfo {volume}
  {149}},\ \bibinfo {pages} {491} (\bibinfo {year} {1966})}\BibitemShut
  {NoStop}%
\bibitem [{\citenamefont {Marcellina}\ \emph {et~al.}(2017)\citenamefont
  {Marcellina}, \citenamefont {Hamilton}, \citenamefont {Winkler},\ and\
  \citenamefont {Culcer}}]{marcellina2017spin}%
  \BibitemOpen
  \bibfield  {author} {\bibinfo {author} {\bibfnamefont {E.}~\bibnamefont
  {Marcellina}}, \bibinfo {author} {\bibfnamefont {A.~R.}\ \bibnamefont
  {Hamilton}}, \bibinfo {author} {\bibfnamefont {R.}~\bibnamefont {Winkler}},\
  and\ \bibinfo {author} {\bibfnamefont {D.}~\bibnamefont {Culcer}},\ }\href
  {https://doi.org/10.1103/PhysRevB.95.075305} {\bibfield  {journal} {\bibinfo
  {journal} {Phys. Rev. B}\ }\textbf {\bibinfo {volume} {95}},\ \bibinfo
  {pages} {075305} (\bibinfo {year} {2017})}\BibitemShut {NoStop}%
\bibitem [{\citenamefont {Rashba}\ and\ \citenamefont
  {Sherman}(1988)}]{rashba1988spin}%
  \BibitemOpen
  \bibfield  {author} {\bibinfo {author} {\bibfnamefont {E.~I.}\ \bibnamefont
  {Rashba}}\ and\ \bibinfo {author} {\bibfnamefont {E.~Y.}\ \bibnamefont
  {Sherman}},\ }\href@noop {} {\bibfield  {journal} {\bibinfo  {journal}
  {Phys.~Lett.~A}\ }\textbf {\bibinfo {volume} {129}},\ \bibinfo {pages} {175}
  (\bibinfo {year} {1988})}\BibitemShut {NoStop}%
\bibitem [{\citenamefont {Marcellina}\ \emph {et~al.}(2018)\citenamefont
  {Marcellina}, \citenamefont {Srinivasan}, \citenamefont {Miserev},
  \citenamefont {Croxall}, \citenamefont {Ritchie}, \citenamefont {Farrer},
  \citenamefont {Sushkov}, \citenamefont {Culcer},\ and\ \citenamefont
  {Hamilton}}]{marcellina2018electrical}%
  \BibitemOpen
  \bibfield  {author} {\bibinfo {author} {\bibfnamefont {E.}~\bibnamefont
  {Marcellina}}, \bibinfo {author} {\bibfnamefont {A.}~\bibnamefont
  {Srinivasan}}, \bibinfo {author} {\bibfnamefont {D.~S.}\ \bibnamefont
  {Miserev}}, \bibinfo {author} {\bibfnamefont {A.~F.}\ \bibnamefont
  {Croxall}}, \bibinfo {author} {\bibfnamefont {D.~A.}\ \bibnamefont
  {Ritchie}}, \bibinfo {author} {\bibfnamefont {I.}~\bibnamefont {Farrer}},
  \bibinfo {author} {\bibfnamefont {O.~P.}\ \bibnamefont {Sushkov}}, \bibinfo
  {author} {\bibfnamefont {D.}~\bibnamefont {Culcer}},\ and\ \bibinfo {author}
  {\bibfnamefont {A.~R.}\ \bibnamefont {Hamilton}},\ }\href
  {https://doi.org/10.1103/PhysRevLett.121.077701} {\bibfield  {journal}
  {\bibinfo  {journal} {Phys. Rev. Lett.}\ }\textbf {\bibinfo {volume} {121}},\
  \bibinfo {pages} {077701} (\bibinfo {year} {2018})}\BibitemShut {NoStop}%
\bibitem [{\citenamefont {Camenzind}\ \emph {et~al.}(2018)\citenamefont
  {Camenzind}, \citenamefont {Yu}, \citenamefont {Stano}, \citenamefont
  {Zimmerman}, \citenamefont {Gossard}, \citenamefont {Loss},\ and\
  \citenamefont {Zumb{\"u}hl}}]{camenzind2018hyperfine}%
  \BibitemOpen
  \bibfield  {author} {\bibinfo {author} {\bibfnamefont {L.~C.}\ \bibnamefont
  {Camenzind}}, \bibinfo {author} {\bibfnamefont {L.}~\bibnamefont {Yu}},
  \bibinfo {author} {\bibfnamefont {P.}~\bibnamefont {Stano}}, \bibinfo
  {author} {\bibfnamefont {J.~D.}\ \bibnamefont {Zimmerman}}, \bibinfo {author}
  {\bibfnamefont {A.~C.}\ \bibnamefont {Gossard}}, \bibinfo {author}
  {\bibfnamefont {D.}~\bibnamefont {Loss}},\ and\ \bibinfo {author}
  {\bibfnamefont {D.~M.}\ \bibnamefont {Zumb{\"u}hl}},\ }\href@noop {}
  {\bibfield  {journal} {\bibinfo  {journal} {Nat.~Commun.}\ }\textbf {\bibinfo
  {volume} {9}},\ \bibinfo {pages} {1} (\bibinfo {year} {2018})}\BibitemShut
  {NoStop}%
\bibitem [{\citenamefont {Bogan}\ \emph {et~al.}(2017)\citenamefont {Bogan},
  \citenamefont {Studenikin}, \citenamefont {Korkusinski}, \citenamefont
  {Aers}, \citenamefont {Gaudreau}, \citenamefont {Zawadzki}, \citenamefont
  {Sachrajda}, \citenamefont {Tracy}, \citenamefont {Reno},\ and\ \citenamefont
  {Hargett}}]{bogan2017consequences}%
  \BibitemOpen
  \bibfield  {author} {\bibinfo {author} {\bibfnamefont {A.}~\bibnamefont
  {Bogan}}, \bibinfo {author} {\bibfnamefont {S.~A.}\ \bibnamefont
  {Studenikin}}, \bibinfo {author} {\bibfnamefont {M.}~\bibnamefont
  {Korkusinski}}, \bibinfo {author} {\bibfnamefont {G.~C.}\ \bibnamefont
  {Aers}}, \bibinfo {author} {\bibfnamefont {L.}~\bibnamefont {Gaudreau}},
  \bibinfo {author} {\bibfnamefont {P.}~\bibnamefont {Zawadzki}}, \bibinfo
  {author} {\bibfnamefont {A.~S.}\ \bibnamefont {Sachrajda}}, \bibinfo {author}
  {\bibfnamefont {L.~A.}\ \bibnamefont {Tracy}}, \bibinfo {author}
  {\bibfnamefont {J.~L.}\ \bibnamefont {Reno}},\ and\ \bibinfo {author}
  {\bibfnamefont {T.~W.}\ \bibnamefont {Hargett}},\ }\href
  {https://doi.org/10.1103/PhysRevLett.118.167701} {\bibfield  {journal}
  {\bibinfo  {journal} {Phys. Rev. Lett.}\ }\textbf {\bibinfo {volume} {118}},\
  \bibinfo {pages} {167701} (\bibinfo {year} {2017})}\BibitemShut {NoStop}%
\bibitem [{\citenamefont {Ekenberg}\ and\ \citenamefont
  {Altarelli}(1984)}]{ekenberg1984calculation}%
  \BibitemOpen
  \bibfield  {author} {\bibinfo {author} {\bibfnamefont {U.}~\bibnamefont
  {Ekenberg}}\ and\ \bibinfo {author} {\bibfnamefont {M.}~\bibnamefont
  {Altarelli}},\ }\href {https://doi.org/10.1103/PhysRevB.30.3569} {\bibfield
  {journal} {\bibinfo  {journal} {Phys. Rev. B}\ }\textbf {\bibinfo {volume}
  {30}},\ \bibinfo {pages} {3569} (\bibinfo {year} {1984})}\BibitemShut
  {NoStop}%
\bibitem [{\citenamefont {Broido}\ and\ \citenamefont
  {Sham}(1985)}]{broido1985effective}%
  \BibitemOpen
  \bibfield  {author} {\bibinfo {author} {\bibfnamefont {D.~A.}\ \bibnamefont
  {Broido}}\ and\ \bibinfo {author} {\bibfnamefont {L.~J.}\ \bibnamefont
  {Sham}},\ }\href {https://doi.org/10.1103/PhysRevB.31.888} {\bibfield
  {journal} {\bibinfo  {journal} {Phys. Rev. B}\ }\textbf {\bibinfo {volume}
  {31}},\ \bibinfo {pages} {888} (\bibinfo {year} {1985})}\BibitemShut
  {NoStop}%
\bibitem [{\citenamefont {Bangert}\ and\ \citenamefont
  {Landwehr}(1985)}]{bangert1985self}%
  \BibitemOpen
  \bibfield  {author} {\bibinfo {author} {\bibfnamefont {E.}~\bibnamefont
  {Bangert}}\ and\ \bibinfo {author} {\bibfnamefont {G.}~\bibnamefont
  {Landwehr}},\ }\href@noop {} {\bibfield  {journal} {\bibinfo  {journal}
  {Superlattice Microst.}\ }\textbf {\bibinfo {volume} {1}},\ \bibinfo {pages}
  {363} (\bibinfo {year} {1985})}\BibitemShut {NoStop}%
\bibitem [{\citenamefont {Goldoni}\ and\ \citenamefont
  {Peeters}(1995)}]{goldoni1995hole}%
  \BibitemOpen
  \bibfield  {author} {\bibinfo {author} {\bibfnamefont {G.}~\bibnamefont
  {Goldoni}}\ and\ \bibinfo {author} {\bibfnamefont {F.~M.}\ \bibnamefont
  {Peeters}},\ }\href {https://doi.org/10.1103/PhysRevB.51.17806} {\bibfield
  {journal} {\bibinfo  {journal} {Phys. Rev. B}\ }\textbf {\bibinfo {volume}
  {51}},\ \bibinfo {pages} {17806} (\bibinfo {year} {1995})}\BibitemShut
  {NoStop}%
\bibitem [{\citenamefont {Wenk}\ \emph {et~al.}(2016)\citenamefont {Wenk},
  \citenamefont {Kammermeier},\ and\ \citenamefont
  {Schliemann}}]{wenk2016conserved}%
  \BibitemOpen
  \bibfield  {author} {\bibinfo {author} {\bibfnamefont {P.}~\bibnamefont
  {Wenk}}, \bibinfo {author} {\bibfnamefont {M.}~\bibnamefont {Kammermeier}},\
  and\ \bibinfo {author} {\bibfnamefont {J.}~\bibnamefont {Schliemann}},\
  }\href@noop {} {\bibfield  {journal} {\bibinfo  {journal} {\prb}\ }\textbf
  {\bibinfo {volume} {93}},\ \bibinfo {pages} {115312} (\bibinfo {year}
  {2016})}\BibitemShut {NoStop}%
\bibitem [{\citenamefont {Dresselhaus}\ \emph {et~al.}(2008)\citenamefont
  {Dresselhaus}, \citenamefont {Dresselhaus},\ and\ \citenamefont
  {Jorio}}]{dresselhaus2008}%
  \BibitemOpen
  \bibfield  {author} {\bibinfo {author} {\bibfnamefont {M.~S.}\ \bibnamefont
  {Dresselhaus}}, \bibinfo {author} {\bibfnamefont {G.}~\bibnamefont
  {Dresselhaus}},\ and\ \bibinfo {author} {\bibfnamefont {A.}~\bibnamefont
  {Jorio}},\ }\href@noop {} {\emph {\bibinfo {title} {Group Theory Application
  to the Physics of Condensed Matter}}}\ (\bibinfo  {publisher} {Springer},\
  \bibinfo {year} {2008})\BibitemShut {NoStop}%
\bibitem [{\citenamefont {Philippopoulos}(2018)}]{philippopoulosgithub}%
  \BibitemOpen
  \bibfield  {author} {\bibinfo {author} {\bibfnamefont {P.}~\bibnamefont
  {Philippopoulos}},\ }\href@noop {} {\bibinfo {title}
  {group-theory-projections}},\ \bibinfo {howpublished}
  {\url{https://github.com/pphili/group-theory-projections}} (\bibinfo {year}
  {2018})\BibitemShut {NoStop}%
\end{thebibliography}%

\end{document}